\documentclass{article}

\usepackage[american]{babel}
\usepackage{bussproofs}
\usepackage{amssymb}
\usepackage{amsthm}
\usepackage{amsmath}

\def\fCenter{{\mbox{\Large$\rightarrow$}}}

\EnableBpAbbreviations

\begin{document}

\newtheorem{definition}{Definition}[section]
\newtheorem{theorem}{Theorem}[section]
\newtheorem{proposition}{Proposition}[section]
\newtheorem{examples}{Examples}[section]
\newtheorem{example}{Example}[section]
\newtheorem{notes}{Remarks}[section]
\newtheorem{note}{Remark}[section]
\newtheorem{notation}{Notation}[section]
\newtheorem{lemma}{Lemma}[section]
\newtheorem{corollary}{Corollary}[section]
\newtheorem{fact}{Fact}[section]
\newtheorem{conjecture}{Conjecture}[section]
\numberwithin{equation}{section}

\title{Proof Search in H\'ajek's Basic Logic}
\author{Simone Bova, Franco Montagna\\[.5ex]
Department of Mathematics and Computer Science\\
University of Siena, Italy\\[.5ex]
\texttt{\{bova,montagna\}@unisi.it}}
\maketitle

\begin{abstract}
We introduce a proof system for H\'ajek's logic \textbf{BL} 
based on a relational hypersequents framework. 
We prove that the rules of our logical calculus, called \textbf{RHBL}, 
are sound and invertible with respect to any valuation of \textbf{BL} into a suitable algebra, called $(\omega)[0,1]$. 
Refining the notion of reduction tree that arises naturally from \textbf{RHBL}, 
we obtain a decision algorithm for \textbf{BL} provability whose running time upper bound is $2^{O(n)}$, 
where $n$ is the number of connectives of the input formula. 
Moreover, if a formula is unprovable, we exploit the constructiveness of a polynomial time algorithm for leaves validity 
for providing a procedure to build countermodels in $(\omega)[0,1]$. 
Finally, since the size of the reduction tree branches is $O(n^3)$, 
we can describe a polynomial time verification algorithm for \textbf{BL} unprovability.
\end{abstract}

\section{Introduction}

A $t$-norm is a binary operation on $[0,1]$ that is 
associative, commutative, weakly increasing and has $1$ as unit. 
Given a $t$-norm $*$, the associated \emph{residuum} is the binary operation $x \rightarrow^* y = max\{  z : z*x \leq y\}$ and 
the associated $t$\emph{-algebra} is the algebra $[0,1]_*=([0,1],*,\rightarrow^*,0)$. 
For each $t$-norm $*$, 
$\text{\bf{L}}^*$ is the propositional logic on the connectives $\odot,\rightarrow$ and the constant $\bot$, 
respectively interpreted on $[0,1]_*$ as $*$, $\rightarrow^*$ and $0$. 
The tautologies of $\text{\bf{L}}^*$ are the formulas evaluating to $1$ on $[0,1]_*$ 
under any valuation of the variables in $[0,1]$. 

In this paper we study the proof-theoretical and proof complexity properties
of a proof search system for H\'ajek's Basic Logic \textbf{BL} \cite{Hajek98a} 
(see Definition~\ref{BL} for \textbf{BL} axioms). 
As shown in \cite{DBLP:journals/soco/CignoliEGT00}, $\textbf{BL}$ is the logic of all continuous $t$-norms and their residua, 
that is, a formula $A$ is a tautology of $\textbf{BL}$ if and only if, for all continuous $t$-norms $*$, $A$ is a tautology of $\text{\bf{L}}^*$. 
Moreover, as shown in \cite{DBLP:journals/apal/BaazH01}, 
the unprovability problem of $\textbf{BL}$ is \textbf{NP}-complete, 
hence the provability problem of $\textbf{BL}$ is decidable.

The proof-theoretical coverage of continuous $t$-norm based logics is not homogeneous. 
On the one hand, 
a variety of logical calculi has been proposed for the fundamental schematic extensions of \textbf{BL}, 
namely \L ukasiewicz logic \textbf{\L}, 
Product logic {\boldmath$\Pi$}, % \cite{DBLP:journals/aml/MetcalfeOG04} and 
G\"odel logic \textbf{G}. 
On the other hand, 
the lack of a natural logical calculus for \textbf{BL} is a recurrent remark in the literature 
\cite{DBLP:conf/lpar/CiabattoniFM04}. 
In \cite{DBLP:conf/ismvl/Aguzzoli04} 
a general method for deriving 
routinely a sequent calculus sound and complete for \textbf{BL} 
(and generally for any continuous $t$-norm based logic) is presented, 
but it seems unsuitable for proof search, 
due to the huge branching factor of the proof trees generated. 
Therefore, as far as we know, the only proof system for \textbf{BL} suitable for proof search 
is the tableaux calculus presented in \cite{DBLP:journals/logcom/MontagnaPT03}, called \textbf{TBL}. 

Unfortunately, we recently discovered a counterexample in \textbf{TBL},
finding that the residuation axiom (A5) of \textbf{BL} is unprovable in \textbf{TBL}, 
or, equivalently, that the leaf of a branch of the \textbf{TBL} reduction tree of (A5) is not an axiom.

For instance, the following branch of the \textbf{TBL} reduction tree of (A5):
\begin{center}
\scriptsize
\begin{prooftree}
\def\extraVskip{3pt}
\def\labelSpacing{2pt}
\def\defaultHypSeparation{\hskip .01in}

\AXC{$\ldots$}

\AXC{$\ldots$}

\AXC{$\ldots$}

\AXC{$\ldots$}

\AXC{$\ldots$}

\AXC{$\ldots$}
\AXC{$ \overline{C \lhd A}, \overline{C \lhd B}, A<\top, B<\top $}
\RightLabel{($\leq$,$ \odot $,$L$)}
\BIC{$ \overline{C < A \odot B}, \overline{C \lhd B}, A<\top, B<\top $}

\RightLabel{($\prec$,$ \odot $,$L$)}
\BIC{$ \overline{C < A \odot B}, \overline{C \lhd B}, ( A \odot B+1+C \prec C+1+A)^+ $}

\RightLabel{($\prec$,$\rightarrow$,$L$)}
\BIC{$ \overline{C < A \odot B}, ( A \odot B + 1 + B\rightarrow C \prec C+1+A )^+$}

\RightLabel{($\prec$,$\rightarrow$,$L$)}
\BIC{$ \overline{C < A \odot B}, ( A\rightarrow(B\rightarrow C) + A \odot B \prec C+1 )^+$}

\RightLabel{($\leq$,$\rightarrow$,$R$)}
\BIC{$ A\rightarrow(B\rightarrow C) \leq (A \odot B)\rightarrow C $}

\RightLabel{($=\top$,$\rightarrow$)}
\BIC{$ ( A\rightarrow(B\rightarrow C) ) \rightarrow ( ( A \odot B )\rightarrow C ) =\top $}

\end{prooftree}
\end{center}
has a leaf that is not an axiom. 
In fact, the leaf corresponds to the disjunction of $\overline{C \lhd A}$, $\overline{C \lhd B}$, $A<\top$ and $B<\top$, 
and its negation, corresponding to the conjunction of $C \lhd A$, $C \lhd B$, $A=\top$ and $B=\top$, 
is satisfied by any valuation of \textbf{BL} into $(\omega)[0,1]$ satisfying $A=B=\top$ e $C<\top$.

Instead of making the necessary changes to \textbf{TBL}, 
we tried to present an alternative and more efficient proof system for \textbf{BL}, 
called \textbf{RWBL}.
From the proof-theoretical point of view,
the kernel of our proof system is a \emph{relational hypersequents} calculus, called \textbf{RHBL},
based on the promising approach that produced 
uniform rules for \textbf{\L}, {\boldmath$\Pi$} and \textbf{G} \cite{DBLP:conf/lpar/CiabattoniFM04}. 
The calculus is \emph{sound} and \emph{complete} and, 
interestingly, it allows for automatic proof search, 
being \emph{cut free} and having the \emph{subformula property}.
From the proof complexity point of view,
we remark that, if a formula $A$ has $n$ connectives,
the proof tree of $A$ in \textbf{RWBL} has height at most $n$.
This fact lowers immediately the upper bounds derivable from \textbf{TBL},
with respect to both the provability and the unprovability problems of \textbf{BL}.
Specifically, the upper bound on the size of a certificate of provability of $A$ is $2^{O(n)}$ 
(compared with the $2^{O(n^2)}$ and $2^{O(n^3)}$ bounds implicit in \cite{DBLP:conf/ismvl/Aguzzoli04} 
and \cite{DBLP:journals/logcom/MontagnaPT03} respectively) 
and the upper bound on the size of a certificate of unprovability of $A$ is $O(n)$ 
as in \cite{DBLP:conf/ismvl/Aguzzoli04} (compared with the $O(n^3)$ bound 
presented in \cite{DBLP:journals/logcom/MontagnaPT03}). 
The time complexity of the algorithms for searching and verifying \textbf{BL} 
proofs follows immediately from these bounds, being dominated by the proof size.

The paper is organized as follows. 
In Section~\ref{preliminariesSection}, we outline the algebraic foundations of our semantic-based proof system. 
In Section~\ref{rhblSection}, we introduce the relational hypersequents calculus \textbf{RHBL} and 
the reduction tree framework \textbf{RWBL}.
In Section~\ref{axiomSection}, we describe a polynomial time algorithm 
for checking the validity of the leaves of a reduction tree. 
In Section~\ref{conpcSection}, we show that our proof system is sound and complete for \textbf{BL} 
and we study its computational complexity.

\section{Preliminaries} \label{preliminariesSection}

\begin{definition}[formula] \label{formulas}
The set of \emph{formulas} is the smallest set F 
such that $\bot \in F$, $\{ p_i : i \in \text{\bf{N}} \} \subseteq F$ 
and, if $A,A' \in F$, then $(A \odot A'),(A \rightarrow A') \in F$. 
The \emph{complexity} $c(A)$ of a formula $A$ 
is the number of connectives occurring in $A$. 
\end{definition}

In the sequel, we write $\top$ for $\bot \rightarrow \bot$ 
and $\circ$ denotes an arbitrary connective. 
Moreover, we say that $B$ is more complex than $A$, $A <_c B$, 
if $c(A)<c(B)$ or $c(A)=c(B)$ but $A \leq_{lex} B$, 
assuming a lexicographical order over the underlying alphabet.

\begin{definition}[BL] \label{BL}
Let $A,B,C \in F$.
The \emph{theory} of \emph{\bf{BL}} is defined by the \emph{modus ponens} deduction rule and the axiom schemata
\begin{enumerate}
\item[\emph{(A1)}] $(A \rightarrow B) \rightarrow ( (B \rightarrow C) \rightarrow (A \rightarrow C) )$
\item[\emph{(A2)}] $(A \odot B) \rightarrow A$
\item[\emph{(A3)}] $(A \odot B) \rightarrow (B \odot A)$
\item[\emph{(A4)}] $(A \odot (A \rightarrow B) ) \rightarrow (B \odot (B \rightarrow A))$
\item[\emph{(A5)}] $( (A \rightarrow (B \rightarrow C)) \leftrightarrow ((A \odot B) \rightarrow C) )$
\item[\emph{(A6)}] $( (A \rightarrow B) \rightarrow C) \rightarrow ( ( (B \rightarrow A) \rightarrow C) \rightarrow C)$
\item[\emph{(A7)}] $\bot \rightarrow A$
\end{enumerate}
A \emph{proof} of a formula $A$ in \emph{\bf{BL}} is a sequence $(A_1, \ldots, A_n)$ of $n$ formulas, $n \in \bf{N}$, 
such that $A_n=A$ and, for all $i = 1, \ldots, n$, $A_i$ is a \emph{\bf{BL}} axiom or
$A_i$ is the result of a \emph{modus ponens} application to formulas $A_j, A_k$, where $j,k < i$.
A formula $A$ is a \emph{\bf{BL}} \emph{theorem} if there is a proof of $A$ in \emph{\bf{BL}} (we write $\emph{\bf{BL}} \vdash A$).
\end{definition}

\begin{definition}[$(\omega)[0,1{]}$ algebra]
The \emph{algebra} $(\omega)[0,1]$ is the algebra on the support $[0,+\infty]$ 
equipped with the operations $*$, $\Rightarrow_*$ and the constants $0$, $+\infty$, 
where $*$ and $\Rightarrow_*$ are respectively defined by
\begin{align*}
x * y & =
\begin{cases}
min(x,y) & \text{if $\lfloor x \rfloor \neq \lfloor y \rfloor$}\\
max(\lfloor x \rfloor,x+y-\lfloor x \rfloor-1) & \text{if $\lfloor x \rfloor=\lfloor y \rfloor<+\infty$}\\
+\infty & \text{if $x=y=+\infty$}
\end{cases} \\
x \Rightarrow_* y & =
\begin{cases}
y & \text{if $\lfloor y \rfloor < \lfloor x \rfloor$}\\
\lfloor x \rfloor+1-x+y & \text{if $\lfloor x \rfloor=\lfloor y \rfloor$ and $y<x$}\\
+\infty & \text{if $x \leq y$}
\end{cases}
\end{align*}
where $\lfloor x \rfloor$ denotes the lower integer part of $x$ and $\lfloor +\infty \rfloor = +\infty$.
\end{definition}

\begin{definition}[valuation] \label{DefinitionValuation} Let $A,B \in F$.
A \emph{valuation} (of \emph{\bf{BL}} into $(\omega)[0,1]$) is a function $v$ such that 
$v(\bot)=0$, $v(p_n) \in [0,+\infty]$, $v(A \odot B)=v(A)*v(B)$ and 
$v(A \rightarrow B)=v(A) \Rightarrow_* v(B)$.
\end{definition}

Let $\overline{v}(A)=v(A)-\lfloor v(A) \rfloor$. 
In the sequel, we say that (the \emph{type} of) a valuation $v$ 
with respect to a pair of formulas $(A,B)$ is: 
$\odot_1$, if $\lfloor v(A) \rfloor < \lfloor v(B) \rfloor$;
$\odot_2$, if $\lfloor v(B) \rfloor < \lfloor v(A) \rfloor$;
$\odot_3$, if $\lfloor v(A) \rfloor = \lfloor v(B) \rfloor <+\infty$ and $1 \leq \overline{v}(A)+\overline{v}(B)$;
$\odot_4$, if $\lfloor v(A) \rfloor =\lfloor v(B) \rfloor <+\infty$ and $\overline{v}(A)+\overline{v}(B) < 1$;
$\odot_5$, if $v(A)=v(B)=+\infty$;
$\rightarrow_1$, if $\lfloor v(B) \rfloor < \lfloor v(A) \rfloor$;
$\rightarrow_2$, if $\lfloor v(A) \rfloor = \lfloor v(B) \rfloor$ and $v(B)<v(A)$;
$\rightarrow_3$, if $v(A) \leq v(B)$.

\begin{fact} \label{PreliminarFact}
Let $A,B \in F$ and $v$ be a valuation of \textbf{BL} into $(\omega)[0,1]$. 
Then, there is exactly one $i \in \{ 1,2,3,4,5\}$ such that $v$ is $\odot_i$ with respect to $(A,B)$ 
and there is exactly one $i \in \{ 1,2,3\}$ such that $v$ is $\rightarrow_i$ with respect to $(A,B)$. 
Moreover, if $\lfloor v(A) \rfloor = \lfloor v(B) \rfloor$, 
then $\lfloor v(A \odot B) \rfloor = \lfloor v(A) \rfloor = \lfloor v(B) \rfloor$, and 
$\lfloor v(A \rightarrow B) \rfloor = \lfloor v(A) \rfloor = \lfloor v(B) \rfloor$ or $\lfloor v(A \rightarrow B) \rfloor = +\infty$.
\end{fact}

The proof system we will introduce in the next section relies on the following characterization of \textbf{BL} provability \cite{DBLP:journals/logcom/MontagnaPT03}.

\begin{theorem} \label{Montagna}
Let $A \in F$. Then, $\emph{\bf{BL}} \vdash A$ if and only if, for every valuation $v$ of \emph{\bf{BL}} into $(\omega)[0,1]$, $v(A)=+\infty$.
\end{theorem}

\section{Logical Rules and Reduction Trees} \label{rhblSection}

\begin{definition}[relational hypersequent] \label{DefinitionRH} 
A \emph{\bf{BL}} \emph{relational sequent} has the form $\Gamma \triangleleft \Delta$, 
where, $\triangleleft \in \{ \ll, \preccurlyeq_z, \prec_z : z \in \text{\bf{Z}} \}$, 
$\Gamma$ and $\Delta$ are finite \emph{multisets} of formulas and, 
if $\triangleleft$ is $\ll$, then $|\Gamma| \leq 1$ and $|\Delta| \leq 1$. 
A \emph{\bf{BL}} \emph{relational hypersequent} is a finite \emph{set} of \emph{\bf{BL}} relational sequents 
of the form $\Gamma_1 \triangleleft_1 \Delta_1 | \ldots | \Gamma_n \triangleleft_n \Delta_n$, 
and $\emptyset$ is the empty relational hypersequent.
The set of \emph{\bf{BL}} relational hypersequents is called RH. 
A relational hypersequent is called \emph{irreducible} if all its formulas are atomic. 
\end{definition}

In the sequel, $\triangleleft \in \{ \ll, \preccurlyeq_z, \prec_z : z \in \text{\bf{Z}} \}$ 
and we write $\triangleleft$ for $\triangleleft_0$.

\begin{definition}[satisfaction, validity] \label{DefinitionSatisfaction} 
Let $G \in RH$ be determined as in Definition~\ref{DefinitionRH} 
and $v$ be a valuation.
Then, $v$ \emph{satisfies} $G$, $v \vDash G$, if, for some $i=1,\ldots,n$:
\begin{enumerate} 
\item[(i)] $\Gamma_i \triangleleft_i \Delta_i$ is $A \ll B$ and $\lfloor v(A) \rfloor < \lfloor v(B) \rfloor$; 

\item[(ii)] $\Gamma_i \triangleleft_i \Delta_i$ is $A \preccurlyeq B$ and both the following conditions hold
\begin{gather}
\lfloor v(A) \rfloor = \lfloor v(B) \rfloor \label{1I} \\
v(A) \leq v(B) \label{1R}
\end{gather} 
and similarly for $\prec$ and $<$ instead of $\preccurlyeq$ and $\leq$ respectively;

\item[(iii)] $\Gamma_i \triangleleft_i \Delta_i$ is $A_1, \ldots, A_n \preccurlyeq_z B_1, \ldots, B_m$, 
where $n,m \geq 0$, $n+m \geq 2$ and, if $n+m = 2$, then $z \neq 0$. 
Moreover, both the following conditions hold
\begin{gather}
\lfloor v(A_1) \rfloor =\ldots= \lfloor v(A_n) \rfloor = \lfloor v(B_1) \rfloor =\ldots= \lfloor v(B_m) \rfloor < +\infty \label{I} \\
\overline{v}(A_1)+\ldots+\overline{v}(A_n)-n \leq \overline{v}(B_1)+\ldots+\overline{v}(B_m)-m+z \label{R}
\end{gather}
and similarly for $\prec$ and $<$ instead of $\preccurlyeq$ and $\leq$ respectively. 
\end{enumerate}

The hypersequent $G$ is \emph{valid} for $\emph{\bf{BL}}$, $\vDash G$, if, 
for all valuations $v$, $v \vDash G$. 
\end{definition}

We remark that condition~(\ref{I}) is stronger than condition~(\ref{1I}), 
since it requires integer parts less than $+\infty$.
We observe that, given a valuation $v$, 
$v$ satisfies $A \ll \top$ only if $v(A)<+\infty$ 
and $v$ satisfies $A \preccurlyeq \top|\top \preccurlyeq A$ only if $v(A)=+\infty$. 
Moreover, $\top \ll A$, $\top \prec A$, $A \prec \top$ and $\emptyset$ are unsatisfiable 
and $\top \prec A|A \prec \top|\top \preccurlyeq A|A \preccurlyeq \top$ is equivalent to $\top \preccurlyeq A$.

Representing the negation of a relational hypersequent $G$ by $\overline{G}$ simplifies the notation of implications. 
For instance, the implication $G \Rightarrow G^\prime$ is represented by $\overline{G}|G^\prime$. 

\begin{notation} \label{NotationAbbreviations}
We introduce the following abbreviations: 
$A \leq B \equiv A \ll B|A \preccurlyeq B$; 
$A \sim B \equiv A \preccurlyeq B|B \preccurlyeq A$; 
$\overline{A \ll B} \equiv A \preccurlyeq B|B \preccurlyeq A|B \ll A$; 
$\overline{A \leq B} \equiv B \prec A|B \ll A$; 
$\overline{A \preccurlyeq B} \equiv A \ll B|B \prec A|B \ll A$; 
$\overline{A \prec B} \equiv A \ll B|B \preccurlyeq A|B \ll A$; 
$\overline{A \sim B} \equiv A \ll B|B \ll A$; 
$\overline{\preccurlyeq_{1} A,B} \equiv \overline{A \sim B}|\overline{A \ll \top}|\overline{B \ll \top}|A,B \prec_{-1}$; 
$\overline{A,B \prec_{-1}} \equiv \overline{A \sim B}|\overline{A \ll \top}|\overline{B \ll \top}|\preccurlyeq_{1} A,B$.
\end{notation}

\begin{fact} \label{PropositionAbbreviationsSoundness}
Let $A,B \in F$ and $v$ be a valuation. 
Then: 
$v \vDash A \leq B$ if and only if $v(A) \leq v(B)$; 
 $v \vDash A \sim B$ if and only if $\lfloor v(A) \rfloor = \lfloor v(B) \rfloor$; 
$v \vDash \overline{A \ll B}$ if and only if $\lfloor v(B) \rfloor \leq \lfloor v(A) \rfloor$; 
$v \vDash \overline{A \leq B}$ if and only if $v(B) < v(A)$; 
$v \vDash \overline{A \preccurlyeq B}$ if and only if $\lfloor v(A) \rfloor \neq \lfloor v(B) \rfloor$ or $v(B) < v(A)$; 
$v \vDash \overline{A \prec B}$ if and only if $\lfloor v(A) \rfloor \neq \lfloor v(B) \rfloor$ or $v(B) \leq v(A)$; 
$v \vDash \overline{A \sim B}$ if and only if $\lfloor v(A) \rfloor \neq \lfloor v(B) \rfloor$; 
$v \vDash \overline{\preccurlyeq_{1} A,B}$ if and only if $\lfloor v(A) \rfloor \neq \lfloor v(B) \rfloor$ or $v(A)=+\infty$ or $v(B)=+\infty$ or $\overline{v}(A)+\overline{v}(B) < 1$; 
$v \vDash \overline{A,B \prec_{-1}}$ if and only if $\lfloor v(A) \rfloor \neq \lfloor v(B) \rfloor$ or $v(A)=+\infty$ or $v(B)=+\infty$ or $1 \leq \overline{v}(A)+\overline{v}(B)$.
\end{fact}

Let $H$ be an arbitrary relational hypersequent, 
$G_{\circ_i}$ be relational hypersequents which will be specified in the sequel, 
$\Gamma, \Delta$ be arbitrary finite multisets of formulas and and $A,B \in F$. 
We define the relational hypersequents calculus $\emph{\bf{RHBL}}$ 
as a set of logical rules having one of the following forms:

{\tiny
\begin{gather}
\mbox{$(\odot,\triangleleft,l)$} \frac{ H|\overline{A \ll B}|G_{\odot_1} \hspace{0.1cm} H|\overline{B \ll A}|G_{\odot_2} \hspace{0.1cm} H|\overline{\preccurlyeq_{1} A,B}|G_{\odot_3} \hspace{0.1cm} H|\overline{A,B \prec_{-1}}|G_{\odot_4} \hspace{0.1cm} H|\overline{\top \preccurlyeq A}|\overline{\top \preccurlyeq B}|G_{\odot_5} }{ H|\Gamma, A \odot B \triangleleft \Delta } \label{FORM1} \\
\mbox{$(\odot,\triangleleft,r)$} \frac{ H|\overline{A \ll B}|G_{\odot_1} \hspace{0.1cm} H|\overline{B \ll A}|G_{\odot_2} \hspace{0.1cm} H|\overline{\preccurlyeq_{1} A,B}|G_{\odot_3} \hspace{0.1cm} H|\overline{A,B \prec_{-1}}|G_{\odot_4} \hspace{0.1cm} H|\overline{\top \preccurlyeq A}|\overline{\top \preccurlyeq B}|G_{\odot_5} }{ H|\Gamma \triangleleft A \odot B,\Delta } \label{FORM2} \\
\mbox{$(\rightarrow,\triangleleft,l)$} \frac{ H|\overline{B \ll A}|G_{\rightarrow_1} \hspace{0.1cm} H|\overline{B \prec A}|G_{\rightarrow_2} \hspace{0.1cm} H|\overline{A \leq B}|G_{\rightarrow_3} }{ H|\Gamma, A \rightarrow B \triangleleft \Delta } \label{FORM3} \\
\mbox{$(\rightarrow,\triangleleft,r)$} \frac{ H|\overline{B \ll A}|G_{\rightarrow_1} \hspace{0.1cm} H|\overline{B \prec A}|G_{\rightarrow_2} \hspace{0.1cm} H|\overline{A \leq B}|G_{\rightarrow_3} }{ H|\Gamma \triangleleft A \rightarrow B,\Delta } \label{FORM4}
\end{gather}}

With respect to (\ref{FORM1})-(\ref{FORM4}) above, we introduce the following terminology. 
The relational hypersequents above are respectively called \emph{premises} (of type) $\circ_1, \ldots, \circ_n$, 
the relational hypersequent below is called \emph{conclusion},  
the formula $A \circ B$ is called \emph{pivot} and 
$H$ is called \emph{side} relational hypersequent of the logical rule. 
With respect to the premise $\circ_i$ of a logical rule, 
the overlined relational hypersequent is called \emph{antecedent} ($\circ_i$) 
and $G_{\circ_i}$ is called \emph{consequent} ($\circ_i$) of the premise.
Notice that, for reasons that will become apparent in Definition~\ref{ReductionTree}, 
we regard the premise of a rule as a relational hypersequents \emph{tuple}, 
ordered by occurrence from left to right.% in the notation of the following definition.

\begin{definition}[RHBL] \label{RHBLLogicalRules}
The relational hypersequent calculus $\emph{\bf{RHBL}}$ 
contains the following rules (we omit the side relational hypersequent of the rules 
and we abbreviate the antecedent of the premise $\circ_i$ with $I_{\circ_i}$): 

{\footnotesize
\begin{gather*}
\mbox{$(\odot,\ll,l)$} \frac{ I_{\odot_1}|A \ll C \hspace{0.3cm} I_{\odot_2}|B \ll C \hspace{0.3cm} I_{\odot_3}|A \ll C \hspace{0.3cm} I_{\odot_4}|A \ll C \hspace{0.3cm} I_{\odot_5} }{ A \odot B \ll C }\\
\mbox{$(\odot,\ll,r)$} \frac{ I_{\odot_1}|C \ll A \hspace{0.3cm} I_{\odot_2}|C \ll B \hspace{0.3cm} I_{\odot_3}|C \ll A \hspace{0.3cm} I_{\odot_4}|C \ll A \hspace{0.3cm} I_{\odot_5}|C \ll \top }{ C \ll A \odot B }\\
\mbox{$(\rightarrow,\ll,l)$} \frac{ I_{\rightarrow_1}|B \ll C \hspace{0.3cm} I_{\rightarrow_2}|A \ll C \hspace{0.3cm} I_{\rightarrow_3} }{ A \rightarrow B \ll C }\\
\mbox{$(\rightarrow,\ll,r)$} \frac{ I_{\rightarrow_1}|C \ll B \hspace{0.3cm} I_{\rightarrow_2}|C \ll A \hspace{0.3cm} I_{\rightarrow_3}|C \ll \top }{ C \ll A \rightarrow B }\\
\mbox{$(\odot,\triangleleft,l)$} \frac{ I_{\odot_1}|\Gamma,A \triangleleft \Delta \hspace{0.3cm} I_{\odot_2}|\Gamma,B \triangleleft \Delta \hspace{0.3cm} I_{\odot_3}|\Gamma,A,B \triangleleft \Delta \hspace{0.3cm} I_{\odot_4}|\Gamma,A,B \triangleleft_{+1} A,B,\Delta  \hspace{0.2cm} I_{\odot_5} }{ \Gamma, A \odot B \triangleleft \Delta }\\
\mbox{$(\odot,\triangleleft,r)$} \frac{ I_{\odot_1}|\Delta \triangleleft A,\Gamma \hspace{0.3cm} I_{\odot_2}|\Delta \triangleleft B,\Gamma \hspace{0.3cm} I_{\odot_3}|\Delta \triangleleft A,B,\Gamma \hspace{0.3cm} I_{\odot_4}|\Delta,A,B \triangleleft_{-1} A,B,\Gamma  \hspace{0.3cm} I_{\odot_5} }{ \Delta \triangleleft A \odot B, \Gamma }\\
\mbox{$(\rightarrow,\triangleleft,l)$} \frac{ I_{\rightarrow_1}|\Gamma,B \triangleleft \Delta \hspace{0.3cm} I_{\rightarrow_2}|\Gamma,B \triangleleft A,\Delta \hspace{0.3cm} I_{\rightarrow_3} }{ \Gamma, A \rightarrow B \triangleleft \Delta }\\
\mbox{$(\rightarrow,\triangleleft,r)$} \frac{ I_{\rightarrow_1}|\Delta \triangleleft B,\Gamma \hspace{0.3cm} I_{\rightarrow_2}|\Delta,A \triangleleft B,\Gamma \hspace{0.3cm} I_{\rightarrow_3} }{ \Delta \triangleleft A \rightarrow B, \Gamma }.
\end{gather*}}

As an exception, if $\Gamma=\emptyset$ and $\Delta=C$, $C \in F$, 
the premise $\odot_5$ of the rules $(\odot,\preccurlyeq,l)$ and $(\odot,\preccurlyeq,r)$ is $I_{\odot_5}|\top \preccurlyeq C$ and 
the premise $\rightarrow_3$ of the rules $(\rightarrow,\preccurlyeq,l)$ and $(\rightarrow,\preccurlyeq,r)$ is $I_{\rightarrow_3}|\top \preccurlyeq C$.
\end{definition}

We point out by the means of an example that 
in the notation of Definition~\ref{RHBLLogicalRules} the side relational hypersequent has been omitted 
and the antecedent of premises has been abbreviated. 
Moreover, we focus on the terminology introduced.

\begin{example}
The fully scripted version of the rule $(\rightarrow,\triangleleft,r)$ is:
{\footnotesize$$\mbox{$(\rightarrow,\triangleleft,r)$} \frac{ H|\overline{B \ll A}|\Delta \triangleleft B,\Gamma \hspace{0.4cm} H|\overline{B \prec A}|\Delta,A \triangleleft B,\Gamma \hspace{0.4cm} H|\overline{A \leq B} }{ H|\Delta \triangleleft A \rightarrow B, \Gamma }$$}
With respect to the rule, 
$H|\Delta \triangleleft A \rightarrow B, \Gamma$ is the conclusion, 
$H$ is the side relational hypersequent, 
$H|\overline{B \ll A}|\Delta \triangleleft B,\Gamma$ is the premise $\rightarrow_1$, 
$H|\overline{B \prec A}|\Delta,A \triangleleft B,\Gamma$ is the premise $\rightarrow_2$ and 
$H|\overline{A \leq B}$ is the premise $\rightarrow_3$. 
With respect to each premise, 
$\overline{B \ll A}$ is the antecedent (of premise) $\rightarrow_1$ and $\Delta \triangleleft B,\Gamma$ is the consequent (of premise) $\rightarrow_1$, 
$\overline{B \prec A}$ is the antecedent (of premise) $\rightarrow_2$ and $\Delta,A \triangleleft B,\Gamma$ is the consequent (of premise) $\rightarrow_2$, 
$\overline{A \leq B}$ is the antecedent (of premise) $\rightarrow_3$ and $\emptyset$ is the consequent (of premise) $\rightarrow_3$.
\end{example}

We remark that some unusual features of the rules, 
namely the emptiness of consequents of premises $\odot_5$ and $\rightarrow_3$ and 
the occurrence of $A,B$ on both sides of $\triangleleft$ in premise $\odot_4$, 
are necessary conditions to prove the semantic properties of the calculus, 
for reasons that will become apparent in the following, 
examining in detail Definition~\ref{DefSound} and Proposition~\ref{PropositionValuationConsequents}.

The properties of soundness and invertibility of a logical rule are defined as follows.

\begin{definition}[soundness, invertibility] \label{DefSound}
Let $R$ be a logical rule. 
Then, $R$ is \emph{sound} for \emph{\bf{BL}} if and only if, 
for all valuations $v$, 
if $v$ satisfies all the premises of $R$, then $v$ satisfies the conclusion of $R$. 
Moreover, $R$ is \emph{invertible} for \emph{\bf{BL}} if and only if, 
for all valuations $v$, 
if $v$ satisfies the conclusion of $R$, then $v$ satisfies all the premises of $R$. 

A logical calculus is sound and invertible if all its rules are sound and invertible.
\end{definition}

We must show that \textbf{RHBL} is sound and invertible. 
To this aim, we anticipate the following two propositions. 

\begin{proposition} \label{PropositionValuationPremises}
Let $R$ be a logical rule determined as in Definition~\ref{RHBLLogicalRules} and $v$ be a valuation. Then, with respect to $(A,B)$:
\begin{flushleft} 
\begin{enumerate} 
\item[(i)] if $v$ is $\odot_i$, then $v$ satisfies the antecedent of the premise $\odot_j$ if and only if $j \neq i$, for all $i,j \in \{1,2,3,4,5\}$;
\item[(ii)] if $v$ is $\rightarrow_i$, then $v$ satisfies the antecedent of the premise $\rightarrow_j$ if and only if $j \neq i$, for all $i,j \in \{1,2,3\}$.
\end{enumerate}
\end{flushleft}

\begin{proof}
Both claims can be proved by cases 
as immediate consequences of Definition~\ref{DefinitionValuation}, Definition~\ref{DefinitionSatisfaction} 
and Fact~\ref{PropositionAbbreviationsSoundness}. 
As an example (the other cases are similar), 
if $v$ is $\odot_3$, then $v \vDash \preccurlyeq_{1} A,B$, that is, 
$v \nvDash \overline{\preccurlyeq_{1} A,B}$ (the antecedent $\odot_3$),
whereas all of $v \vDash \overline{A \ll B}$ (the antecedent $\odot_1$), $v \vDash \overline{B \ll A}$ (the antecedent $\odot_2$),
$v \vDash \overline{A,B \prec_{-1}}$ (the antecedent $\odot_4$) and $v \vDash \overline{\top \preccurlyeq A}|\overline{\top \preccurlyeq B}$ (the antecedent $\odot_5$) hold,  
thus the claim holds in this case.
\end{proof}
\end{proposition}

\begin{proposition} \label{PropositionValuationConsequents}
Let $R$ be a logical rule determined as in Definition~\ref{RHBLLogicalRules} and $v$ be a valuation. Then, with respect to $(A,B)$:
\begin{flushleft} 
\begin{enumerate} 
\item[(i)] if $v$ is $\odot_i$, then $v$ satisfies the consequent of premise $\odot_i$ if and only if $v$ satisfies the conclusion, for all $i \in \{1,2,3,4,5\}$;
\item[(ii)] if $v$ is $\rightarrow_i$, then $v$ satisfies the consequent of premise $\rightarrow_i$ if and only if $v$ satisfies the conclusion, for all $i \in \{1,2,3\}$;
\end{enumerate}
\end{flushleft}

\begin{proof} The argument relies on Definitions \ref{DefinitionValuation} and \ref{DefinitionSatisfaction} and on Fact~\ref{PreliminarFact}. 

($i$) To prove the first claim we examine the $\odot$ rules. 
Consider ($\odot,\ll,l$) and ($\odot,\ll,r$). 
If $v$ is $\odot_1,\odot_3,\odot_4$, 
then $\lfloor v(A \odot B) \rfloor=\lfloor v(A) \rfloor$, 
thus $v$ satisfies the consequent of premise $\odot_i$ if and only if $v$ satisfies the conclusion, for $i=1,3,4$. 
If $v$ is $\odot_2$, $\lfloor v(A \odot B) \rfloor=\lfloor v(B) \rfloor$, 
thus $v$ satisfies the consequent of premise $\odot_2$ if and only if $v$ satisfies the conclusion. 
If $v$ is $\odot_5$, the consequent of premise $\odot_5$ is equivalent to the conclusion, 
since, for the left version of the rule, $\emptyset$ and $\top \ll C$ are both unsatisfiable, 
and, for the right version of the rule, $v$ satisfies $C \ll \top$ if and only if $v$ satisfies the conclusion.

Consider ($\odot,\triangleleft,l$) and ($\odot,\triangleleft,r$). 
If $v$ is $\odot_1$ ($\odot_2$), then $v$ satisfies the consequent of premise $\odot_1$ ($\odot_2$) if and only if $v$ satisfies the conclusion. 
If $v$ is $\odot_3$, first we observe that
\begin{equation}
\begin{split}
\overline{v}(A \odot B)-1 & =v(A)+v(B)-\lfloor v(A) \rfloor-1-\lfloor v(B) \rfloor-1 \\
                          & =\overline{v}(A)-1+\overline{v}(B)-1 \label{C3}
\end{split}
\end{equation}
Otherwise stated, the value of a relation of the form $\Gamma,A \odot B \triangleleft \Delta$ 
does not change replacing $A \odot B$ on the left with $A,B$ and, 
similarly, the value of a relation of the form $\Gamma \triangleleft A \odot B,\Delta$ 
does not change replacing $A \odot B$ on the right with $A,B$. 
Then we consider two cases, the first arising when $|\Gamma| > 0$ or $|\Delta| > 1$ 
and the second arising when $|\Gamma|=0$ and $|\Delta| = 1$, say, $\Delta=C$ for some $C \in F$.
In the first case, $v$ satisfies conditions (\ref{I}) and (\ref{R}) in the consequent of premise $\odot_3$ 
if and only if $v$ satisfies conditions (\ref{I}) and (\ref{R}) in the conclusion, 
exploiting (\ref{C3}) and Fact~\ref{PreliminarFact}. 
In the second case, 
for the soundness, if $v$ satisfies the consequent of premise $\odot_3$, 
then, by condition (\ref{I}) of Definition~\ref{DefinitionSatisfaction}($iii$), $\lfloor v(A) \rfloor = \lfloor v(B) \rfloor= \lfloor v(C) \rfloor <+\infty$, 
thus also $\lfloor v(C) \rfloor = \lfloor v(A \odot B) \rfloor$ (weakening condition (\ref{I})), 
and $v$ satisfies condition (\ref{1I}) in the conclusion, 
but $v$ satisfies also condition (\ref{1R}), by (\ref{C3}), 
thus $v$ satisfies the conclusion. 
For the invertibility, if $v$ satisfies the conclusion, then, by condition (\ref{1I}) of Definition~\ref{DefinitionSatisfaction}($ii$), 
$\lfloor v(C) \rfloor = \lfloor v(A \odot B) \rfloor$, 
then $\lfloor v(A) \rfloor = \lfloor v(B) \rfloor= \lfloor v(C) \rfloor <+\infty$, 
thus $v$ satisfies condition (\ref{I}) of Definition~\ref{DefinitionSatisfaction}($iii$) in the consequent of premise $\odot_3$, 
but $v$ satisfies also the condition (\ref{R}), by (\ref{C3}), 
thus $v$ satisfies the consequent of premise $\odot_3$. 
Notice that we rely on the assumption that $v$ is $\odot_3$ for strengthening condition (\ref{1I}).
If $v$ is $\odot_4$, first we observe that
\begin{equation}
\begin{split}
\gamma+\overline{v}(A \odot B)-1 \leq \delta & \Leftrightarrow \gamma + \lfloor v(A) \rfloor -\lfloor v(A) \rfloor -1 \leq \delta \\
                                             & \Leftrightarrow \gamma \leq \delta+1 \label{C41} \\
\end{split}
\end{equation}
$\gamma, \delta \in \text{\bf{R}}^+$. Otherwise stated, 
the value of a relation $\Gamma,A \odot B \triangleleft \Delta$ does not change 
replacing $A \odot B$ on the left with $A,B$, replacing $\Delta$ on the right with $A,B,\Delta$ 
and increasing the integer index of the relation by 1. Moreover, we observe that
\begin{equation}
\begin{split}
\gamma \leq \overline{v}(A \odot B)-1+\delta & \Leftrightarrow \gamma \leq \lfloor v(A) \rfloor -\lfloor v(A) \rfloor -1 +\delta\\
                                             & \Leftrightarrow \gamma \leq \delta-1 \label{C42} \\
\end{split}
\end{equation}
$\gamma, \delta \in \text{\bf{R}}^+$. Otherwise stated, 
the value of a relation $\Gamma \triangleleft A \odot B,\Delta$ does not change 
replacing $\Gamma$ on the left with $\Gamma,A,B$, replacing $A \odot B$ on the right with $A,B$ 
and decreasing the integer index of the relation by 1. 
Then, treating the two cases mentioned above as for $\odot_3$ valuations and exploiting (\ref{C41}) and (\ref{C42}), 
we obtain the equivalence modulo $v$ of the consequent of premise $\odot_4$ and the conclusion. 
It may be worth mentioning that the apparent redundancy of $A,B$ on both sides of the consequent of premise $\odot_4$ is necessary to obtain the soundness of the rule. 
The problem is to guarantee the condition (\ref{I}) in the conclusion, 
whereas to guarantee condition (\ref{R}) a premise $\Gamma \triangleleft \Delta$ would be sufficient. 
In fact, it is clear that the value of a relation $\Gamma,A,B \triangleleft A,B,\Delta$ 
is equal to the value of the relation $\Gamma \triangleleft \Delta$. 
If $v$ is $\odot_5$, we consider the two cases mentioned above. In the first case, 
neither the consequent of premise $\odot_5$ nor the conclusion is satisfied by $v$, 
the former being unsatisfiable and the latter violating condition (\ref{I}) of Definition~\ref{DefinitionSatisfaction}($iii$). 
In the second case, we distinguish three subcases. 
If $\triangleleft$ is $\prec_z$, 
neither the consequent of premise $\odot_5$ nor the conclusion is satisfied by $v$, 
the former being unsatisfiable and the latter violating condition (\ref{R}) of Definition~\ref{DefinitionSatisfaction}($ii$). 
If $\triangleleft$ is $\preccurlyeq_z$ and $z=0$, 
then $v$ satisfies the consequent of premise $\odot_5$ if and only if $\Delta=\top$ if and only if $v$ satisfies also the conclusion, 
by the exception outlined in Definition~\ref{RHBLLogicalRules}. 
If $\triangleleft$ is $\preccurlyeq_z$ and $z \neq 0$, 
neither the consequent of premise $\odot_5$ (which is $\emptyset$) nor the conclusion is satisfied by $v$, 
the former being unsatisfiable and the latter violating condition (\ref{I}) of Definition~\ref{DefinitionSatisfaction}($iii$), 
since $\lfloor A \odot B \rfloor = + \infty$.

($ii$) The proof is entirely analogous to ($i$). As a hint, we observe that
\begin{equation}
\begin{split}
\overline{v}(A \rightarrow B)-1 \leq 0 & \Leftrightarrow \lfloor v(A) \rfloor +1 -v(A) +v(B) -\lfloor v(B) \rfloor-1 \leq 0 \\
                                       & \Leftrightarrow v(B)-\lfloor v(B) \rfloor-1 \leq v(A)-\lfloor v(A) \rfloor-1 \\
                                       & \Leftrightarrow \overline{v}(B)-1 \leq \overline{v}(A)-1 \label{I21} \\
\end{split}
\end{equation}
Otherwise stated, the value of a relation $\Gamma,A \rightarrow B \triangleleft \Delta$ does not change
replacing $A \rightarrow B$ on the left with $B$ and replacing $\Delta$ on the right with $A,\Delta$.
Moreover, we observe that
\begin{equation}
\begin{split}
0 \leq \overline{v}(A \rightarrow B)-1 & \Leftrightarrow 0 \leq \lfloor v(A) \rfloor +1 -v(A) +v(B) -\lfloor v(B) \rfloor-1 \\
                                       & \Leftrightarrow v(A)-\lfloor v(A) \rfloor-1 \leq v(B)-\lfloor v(B) \rfloor-1 \\
                                       & \Leftrightarrow \overline{v}(A)-1 \leq \overline{v}(B)-1 \label{I22} \\
\end{split}
\end{equation}
Otherwise stated, the value of a relation $\Delta \triangleleft A \rightarrow B,\Gamma$ does not change
replacing $\Delta$ on the left with $\Delta,A$ and replacing $A \rightarrow B$ on the right with $B$.

\end{proof}
\end{proposition}

\begin{lemma} \label{CalculusSoundness}
The calculus \emph{\bf{RHBL}} is sound and invertible.
\begin{proof} 
Let the logical rules for $\odot$ and $\rightarrow$ be determined as in Definition~\ref{RHBLLogicalRules} 
and let $v$ be a valuation.

We prove that $\odot$ rules are sound and invertible. 
For soundness, by Fact~\ref{PreliminarFact}, $v$ is $\odot_i$ for exactly one $i \in \{ 1,\ldots,5 \}$ 
and, by Proposition~\ref{PropositionValuationPremises}($i$), if $v$ is $\odot_i$, then $v$ satisfies the antecedent of a premise $\odot_j$ only if $j \neq i$. 
If $v$ does not satisfy the premise $\odot_i$, the claim holds trivially. Otherwise, if $v$ satisfies the premise $\odot_i$, 
then, by Proposition~\ref{PropositionValuationPremises}($i$), $v$ must satisfy the consequent of the premise $\odot_i$, 
then, by Proposition~\ref{PropositionValuationConsequents}($i$), $v$ satisfies the conclusion of the rule, 
and the claim holds. 
For invertibility, if $v$ does not satisfy the conclusion, the claim holds trivially. 
Otherwise, let $v$ be a $\odot_i$ valuation such that $v$ satisfies the conclusion. 
By Proposition~\ref{PropositionValuationConsequents}($i$), $v$ satisfies (the consequent of) the premise $\odot_i$ 
and, by Proposition~\ref{PropositionValuationPremises}($i$), $v$ satisfies (the antecedent of) the premises $\odot_j$, for all $j \neq i$, 
thus the claim holds.

The argument for proving soundness and invertibility of $\rightarrow$ rules is similar.
\end{proof}
\end{lemma}

In the remainder of this section we will introduce and refine a notion of \emph{reduction} based on \textbf{RHBL}, 
which will form the kernel of a decision algorithm for the validity problem of \textbf{BL} 
(see Section~\ref{conpcSection}). 

\begin{definition}[RHBL reduction] \label{ReductionTree}
A \emph{\bf{RHBL}} \emph{reduction} of $A \in F$ is a labeled rooted ordered tree $T_A$ such that all the following statements hold.
\begin{enumerate}
\item[(i)] The \emph{root} of the tree is labeled by the relational hypersequent $\top \preccurlyeq A$.
\item[(ii)] If a node $u$ of the tree is labeled by an irreducible relational hypersequent, then $u$ is a \emph{leaf} of the tree.
\item[(iii)] Let $u$ be a node of the tree labeled by a reducible relational hypersequent $G$, 
let $A \circ A^\prime$ be the most complex formula of $G$ with respect to $<_c$ and 
choose an arbitrary occurrence of $A \circ A^\prime$ in a relational context of type $(\triangleleft,s)$, $s \in \{ l, r \}$. 
Then the proper descendants of $u$ are labeled by the premises of the logical rule $(\circ,\triangleleft,s)$ 
with conclusion $G$ and pivot $A \circ A^\prime$, 
the $i$th child being labeled by the premise $\circ_i$.
\end{enumerate}
The length of the path from the root to a node $u$ is the \emph{depth} of the node, which we write $d(u)$. 
A simple downward path from the root to a leaf is a \emph{branch} of the tree. 
The \emph{height} of a tree $T$, $h(T)$, is the number of edges on the longest branch.
\end{definition}

The tree produced by Definition~\ref{ReductionTree} allows for refinements. Consider the following example.

\begin{example}  \label{Reduction1}
Let $G_\rightarrow=H|A \rightarrow B \ll \Delta_1|\Delta_2 \prec A \rightarrow B,\Gamma_2$, 
where $A \rightarrow B$ is the most complex formula of $G_\rightarrow$ and $A \rightarrow B$ does not occur in $H$. 
We consider the portion of the \emph{\bf{RHBL}} reduction tree which is rooted at $G_\rightarrow$ and 
follows the second child of the nodes, corresponding to $\rightarrow_2$ premises. The reduction is:
\begin{center}
\scriptsize
\begin{prooftree}
\def\extraVskip{3pt}
\def\labelSpacing{2pt}
\def\defaultHypSeparation{\hskip .01in}  
  \AXC{$\ldots$}
    \AXC{$\ldots$}
    \AXC{$H|\overline{B \prec A}|A \ll \Delta_1|\Delta_2,A \prec B,\Gamma_2$}
    \AXC{$\ldots$}
  \RightLabel{$(\rightarrow,\prec,l)$}
  \TIC{$H|\overline{B \prec A}|A \ll \Delta_1|\Delta_2,\prec A \rightarrow B,\Gamma_2$}
  \AXC{$\ldots$}
\RightLabel{$(\rightarrow,\ll,l)$}
\TIC{$H|A \rightarrow B \ll \Delta_1|\Delta_2 \prec A \rightarrow B,\Gamma_2$}
\end{prooftree}
\end{center}
Notice that the antecedent $\rightarrow_2$ is sufficient to eliminate both occurrences of $A \rightarrow B$ 
and that, in this case, eliminating two occurrences of $A \rightarrow B$ requires two steps. 
\end{example}

In the example, given a relational hypersequent $G$ having $A \circ B$ as its most complex formula with respect to $<_c$, 
the branches rooted at $G$ start eliminating each occurrence of $A \circ B$, 
since the pivot of the rule is chosen as the most complex formula of $G$ by definition. 
To this aim, the very same antecedent is used repeatedly along the branch to eliminate each occurrence of $A \circ B$ until all occurrences are eliminated. 
This scenario suggests the idea to exploit the antecedent for eliminating \emph{all} the occurrences of the most complex formula via a concise rule.

For this purpose, we introduce \emph{substitutions} on relational hypersequents, 
relying on the usual multiset operations for the manipulation of the formulas. 
Let $G,G^{'} \in RH$ and $A,B,B^{'} \in F$, 
then:
\begin{enumerate}
\item[$(i)$] the relational hypersequent $G(A \Leftarrow B)$ is obtained replacing each occurrence of $A$ with $B$ in $G$; 
\item[$(ii)$] the relational hypersequent $G(A \Leftarrow B,B^{'})$ is obtained replacing each occurrence of $A$ with $B,B^{'}$ in $G$; 
\item[$(iii)$] if the occurrences of $A \odot B$ on the left and right of $\triangleleft$ in $G$ are $l$ and $r$ respectively, 
then the relational hypersequent $G(A \odot B \Leftarrow A,B \triangleleft_{+l-r} A,B)$ 
is obtained deleting \emph{all} the occurrences of $A \odot B$ in relations of type $\triangleleft$ in $G$, 
adding \emph{one} copy of $A,B$ on the left and right of $\triangleleft$ and 
updating by $l-r$ the index of $\triangleleft$; 
\item[$(iv)$] if the occurrences of $A \rightarrow B$ on the left and right of $\triangleleft$ in $G$ are $l$ and $r$ respectively, 
then the relational hypersequent $G(A \rightarrow B \Leftarrow A^r,B^l \triangleleft A^l,B^r)$ 
is obtained deleting all the occurrences of $A \rightarrow B$ in relations of type $\triangleleft$ in $G$, 
adding $r$ occurrences of $A$ and $l$ occurrences of $B$ on the left of $\triangleleft$ and 
adding $l$ occurrences of $B$ and $r$ occurrences of $A$ on the right of $\triangleleft$. 
\end{enumerate}
Moreover, if $\sigma$ and $\sigma^{'}$ are substitutions as above, 
then the relational hypersequent $G\sigma|G^{'}\sigma^{'}$ 
is obtained executing simultaneously $\sigma$ over $G$ and $\sigma^{'}$ over $G^{'}$. 

The notion of rewriting rule is based upon the substitution operation 
and provides an effective means for eliminating all occurrences of a formula from a relational hypersequent.

\begin{definition}[rewriting rules] \label{RHBLRewritingRules}
Let $H, G_{\circ},G_{\circ}^{'},G_{\circ}^{''},G_{\circ_i}^{'''} \in RH$ 
such that $A \circ B$ is the most complex formula occurring in $G_{\circ}$. 
Moreover, assume that $H \subseteq G_{\circ}$ is the set of relational sequents in which $A \circ B$ does not occur, 
$G_{\circ}^{'} \subseteq G_{\circ}$ is the set of $\ll$ relational sequents in which $A \circ B$ occurs, 
that $G_{\circ}^{''} \subseteq G_{\circ}$ is the set of $\prec_z,\preccurlyeq_z$ relational sequents in which $A \circ B$ occurs 
and that $G_{\circ}^{'''} \subseteq G_{\circ}^{''}$ is the set of $\preccurlyeq$ relational sequents with at most one formula on each side. 

The \emph{\bf{BL}} \emph{rewriting rules} for $\odot$ and $\rightarrow$ are:

{\footnotesize
\begin{gather*}
\mbox{$(\odot)$} \frac{ \overline{A \ll B}|G_{\odot_1} \hspace{0.1cm} \overline{B \ll A}|G_{\odot_2} \hspace{0.1cm} \overline{\preccurlyeq_{1} A,B}|G_{\odot_3} \hspace{0.1cm} \overline{A,B \prec_{-1}}|G_{\odot_4} \hspace{0.1cm} \overline{\top \preccurlyeq A}|\overline{\top \preccurlyeq B}|G_{\odot_5} }{ G_{\odot} }\\
\mbox{$(\rightarrow)$} \frac{ \overline{B \ll A}|G_{\rightarrow_1} \hspace{0.1cm} \overline{B \prec A}|G_{\rightarrow_2} \hspace{0.1cm} \overline{A \leq B}|G_{\rightarrow_3} }{ G_{\rightarrow} }
\end{gather*}}
where $G_{\circ_i}$ is obtained by the following substitutions of $A \circ B$ in $G_{\circ}$ 
($C=A$ if $c(A) <_c c(B)$ and $C=B$ otherwise):
\begin{flushleft} 
\begin{enumerate}
\item[(i)] $G_{\odot_1}=G_{\odot}(A \odot B \Leftarrow A)|H$;
\item[(ii)] $G_{\odot_2}=G_{\odot}(A \odot B \Leftarrow B)|H$;
\item[(iii)] $G_{\odot_3}=G_{\odot}^{'}(A \odot B \Leftarrow C)|G_{\odot}^{''}(A \odot B \Leftarrow A,B)|H$;
\item[(iv)] $G_{\odot_4}=G_{\odot}^{'}(A \odot B \Leftarrow C)|G_{\odot}^{''}(A \odot B \Leftarrow A,B \triangleleft_{+l-r} A,B)|H$;
\item[(v)] $G_{\odot_5}=G_{\odot}^{'}|G_{\odot}^{'''}(A \odot B \Leftarrow \top)|H$;
\item[(vi)] $G_{\rightarrow_1}=G_{\rightarrow}(A \rightarrow B \Leftarrow B)|H$;
\item[(vii)] $G_{\rightarrow_2}=G_{\rightarrow}^{'}(A \rightarrow B \Leftarrow C)|G_{\rightarrow}^{''}(A \rightarrow B \Leftarrow A^r,B^l \triangleleft A^l,B^r)|H$;
\item[(viii)] $G_{\rightarrow_3}=G_{\rightarrow}^{'}|G_{\rightarrow}^{'''}(A \rightarrow B \Leftarrow \top)|H$.
\end{enumerate}
\end{flushleft}

With respect to a rewriting rule, \emph{premise} (of type) $\circ_i$, \emph{conclusion}, \emph{side}, \emph{antecedent} (of type) $\circ_i$ 
are defined as in Definition~\ref{RHBLLogicalRules}. 
The \emph{consequent} (of type) $\circ_i$ is $G_{\circ_i}$ and the \emph{pivot} is $A \circ B$.
\end{definition}

We observe that, if $n$ is the number of distinct subformulas occurring in the conclusion of a rule, 
then the number of distinct subformulas occurring in each premise of the rule is at most $n-1$, 
since the pivot occurs in the conclusion but not in the premises. 

The semantic properties required follow immediately from the soundness and invertibility of logical rules.

\begin{corollary} \label{RewritingSoundness}
The rewriting rules for \emph{\bf{BL}} are sound and invertible.
\begin{proof}
The argument used to prove soundness and invertibility of logical rules in Proposition~\ref{CalculusSoundness} works equally well for rewriting rules, 
once proved that, for both rewriting rules, the consequent of premise $\circ_i$ and the conclusion are equivalent 
modulo a valuation $v$ of type $\circ_i$. 
But this claim follows from Fact~\ref{PreliminarFact}, 
from (\ref{C3}), (\ref{C41}), (\ref{C42}), (\ref{I21}), (\ref{I22}) in the proof of Proposition~\ref{PropositionValuationConsequents} 
and from the definition of the substitutions and Definition~\ref{RHBLRewritingRules}.
\end{proof}
\end{corollary}

It is easy to realize that, with respect to the construction of the reduction tree, 
the effect of a unique rewriting rule having pivot $A$ on a given relational hypersequent is equivalent to 
the effect of a sequence of logical rule having the same pivot. 
Consider the following example.

\begin{example} \label{Substitution1}
Let $G_\rightarrow$ be as in Example~\ref{Reduction1}. 
In the setting of Definition~\ref{RHBLRewritingRules} we have:
{\small
\begin{equation*}
\begin{split}
%G_\rightarrow & = H|A \rightarrow B \ll \Delta_1|\Delta_2 \prec A \rightarrow B,\Gamma_2\\
G_{\rightarrow}^{'} & = A \rightarrow B \ll \Delta_1\\
G_{\rightarrow}^{''} & =\Delta_2 \prec A \rightarrow B,\Gamma_2\\
G_{\rightarrow_2} & = G_{\rightarrow}^{'}( \Leftarrow A )|G_{\rightarrow}^{''}(A \rightarrow B \Leftarrow A^1,B^0 \prec A^0,B^1)\\
                  & = A \rightarrow B \ll \Delta_1( \Leftarrow A )|\Delta_2 \prec A \rightarrow B,\Gamma_2(A \rightarrow B \Leftarrow A^1,B^0 \prec A^0,B^1) \\
                  & = A \ll \Delta_1|\Delta_2,A \prec B,\Gamma_2
\end{split}
\end{equation*}}
The rewriting rule $(\rightarrow)$ produces the relational hypersequent $H|\overline{B \prec A}|A \ll \Delta_1|\Delta_2,A \prec B,\Gamma_2$ in one step, 
whereas the logical rules of Example~\ref{Reduction1} require two steps to obtain the same relational hypersequent.
\end{example}

Comparing Example~\ref{Reduction1} and \ref{Substitution1}, we observe that the rewriting rules require 1 step constant to eliminate all occurrences of a formula, 
whereas in the worst case the logical rules require $n$ steps to achieve the objective, if the formula occurs $n$ times. 
In this light, a refined notion of reduction arises naturally.

\begin{definition}[RWBL reduction] \label{RewritingTree}
The \emph{\bf{RWBL}} \emph{reduction} of $A \in F$ is a labeled rooted ordered tree $T_A$ such that all the following statements hold.
\begin{enumerate}
\item[(i)] The \emph{root} of the tree is labeled by the relational hypersequent $\top \preccurlyeq A$.
\item[(ii)] If a node $u$ of the tree is labeled by an irreducible relational hypersequent, then $u$ is a \emph{leaf} of the tree.
\item[(iii)] Let $u$ be a node of the tree labeled by a reducible relational hypersequent $G$ and $A \circ A^\prime$ be the most complex formula of $G$. 
Then the proper descendants of $u$ are labeled by the premises of the rewriting rule $(\circ)$ with conclusion $G$ and pivot $A \circ A^\prime$, 
the $i$th child being labeled by the premise $\circ_i$.
\end{enumerate}
\emph{Depth}, \emph{branch}, \emph{height} and \emph{size} are defined as in Definition~\ref{ReductionTree}
\end{definition}

Notice that the rewriting rule $(\odot)$ handles $\odot_4$ premise 
avoiding the combinatorial explosion of formulas along the branches of the reduction tree. 
This complexity refinement of logical rules explains why 
the irreducible relational hypersequents produced by \textbf{RHBL} and \textbf{RWBL} reductions, however equivalent, 
are distinct. Inspect the following example.

\begin{example} \label{Substitution0}
Let $G=\Delta_1,A \odot B,A \odot B \preccurlyeq_z A \odot B,\Gamma_1$, 
where $A \odot B$ is the most complex formula of $G$. 
We consider the portion of the \emph{\bf{RHBL}} and \emph{\bf{RWBL}} reductions tree which are rooted at $G$ and 
follows the fourth child of the nodes, corresponding to $\odot_4$ premises. The \emph{\bf{RHBL}} reduction is:
\begin{center}
\footnotesize
\begin{prooftree}
\def\extraVskip{3pt}
\def\labelSpacing{2pt}
\def\defaultHypSeparation{\hskip .01in}
	\AXC{$\ldots$}
		\AXC{$\ldots$}
	        \AXC{$\ldots$}		
			\AXC{$\overline{A,B \prec_{-1}}|A^3,B^3 \triangleleft_{z+1} A^3,B^3$}
			\AXC{$\ldots$}
		\RightLabel{$(\odot,\triangleleft_{z+2},r)$}
		\TIC{$\overline{A,B \prec_{-1}}|A^2,B^2 \triangleleft_{z+2} A^2,B^2,A \odot B$}
		
		\AXC{$\ldots$}
	\RightLabel{$(\odot,\triangleleft_{z+1},l)$}
	\TIC{$\overline{A,B \prec_{-1}}|A \odot B,A,B \triangleleft_{z+1} A,B,A \odot B$}
	\AXC{$\ldots$}
\RightLabel{$(\odot,\triangleleft_z,l)$}
\TIC{$A \odot B^2 \triangleleft_z A \odot B$}
\end{prooftree}
\end{center}
and the \emph{\bf{RWBL}} reduction is:
\begin{center}
\footnotesize
\begin{prooftree}
\def\extraVskip{3pt}
\def\labelSpacing{2pt}
\def\defaultHypSeparation{\hskip .01in}
	\AXC{$\ldots$}
	\AXC{$\overline{A,B \prec_{-1}}|A,B \triangleleft_{z+1} A,B$}
	\AXC{$\ldots$}
\RightLabel{$(\odot)$}
\TIC{$A \odot B^2 \triangleleft_z A \odot B$}
\end{prooftree}
\end{center}
As in Example~\ref{Substitution1}, the \emph{\bf{RWBL}} reduction requires one step to eliminate $A \odot B$, 
whereas the \emph{\bf{RHBL}} reduction requires one step for each occurrence of $A \odot B$. 
Moreover, the irreducible relational hypersequent obtained via \emph{\bf{RWBL}} reduction is smaller 
(but equivalent, as can be easily checked applying Definition~\ref{DefinitionSatisfaction}).
\end{example}

We will take advantage of this remarkable feature of \emph{\bf{RWBL}} while studying the size of the \emph{\bf{RWBL}} reductions. 
By means of the example below, we illustrate another behaviour of the \emph{\bf{RWBL}} reductions, 
which will be exploited, again in Lemma~\ref{Complexity}, to avoid the size explosion of the tree branches.

\begin{example} \label{Substitution2}
We consider the following portion of a \emph{\bf{RWBL}} reduction tree, 
generated applying upwards the $\odot$ rewriting rule 
and showing only the fourth child of the nodes 
(that is, the $\odot_4$ premise of the rule):
\begin{center}
\footnotesize
\begin{prooftree}
\def\extraVskip{3pt}
\def\labelSpacing{2pt}
\def\defaultHypSeparation{\hskip .01in}
	\AXC{$\ldots$}
		\AXC{$\ldots$}
	        \AXC{$\ldots$}		
			\AXC{$\overline{C,D \prec_{-1}}|A,B,C,D \triangleleft_{+1} A,B,C,D,\Delta$}
			\AXC{$\ldots$}
		\RightLabel{$(\odot)$}
		\TIC{$\overline{A,B \prec_{-1}}|A,B,C \odot D \triangleleft_{+1} A,B,C \odot D,\Delta$}
		
		\AXC{$\ldots$}
	\RightLabel{$(\odot)$}
	\TIC{$\overline{A \odot B,C \odot D \prec_{-1}}|A \odot B,C \odot D \triangleleft_{+1} A \odot B,C \odot D,\Delta$}
	\AXC{$\ldots$}
\RightLabel{$(\odot)$}
\TIC{$(A \odot B) \odot (C \odot D) \triangleleft \Delta$}
\end{prooftree}
\end{center}
Once $(A \odot B) \odot (C \odot D)$, which occurs only on the left, is eliminated, 
the consequent of premise $\odot_4$ contains the pair $A \odot B,C \odot D$ on both sides, 
thus its size is almost two times the size of the conclusion. 
But the subsequent elimination of $A \odot B$ takes advantage of its balanced occurrence, 
since replacing $A \odot B$ with the pair $A,B$ on both sides does not affect the size of the result 
(the same argument holds for $C \odot D$). 
\end{example}

It is easy to realize that the phenomenon described is always verified, relying on Definition~\ref{RHBLRewritingRules}. 
Lemma~\ref{Complexity} will make this argument rigorous.

The following proposition establishes two structural properties of \textbf{RWBL} reductions. 

\begin{proposition} \label{LeafProperty}
Let $T_A$ be a \emph{\bf{RWBL}} reduction of a formula $A$ and $G$ be a leaf of $T_A$.
\begin{enumerate}
\item[(i)] Let $q_1,\dots,q_s \triangleleft q_{s+1},\dots,q_t$ be a relational sequent occurring in $G$, 
$\triangleleft \in \{ \prec_z, \preccurlyeq_z \}$, 
$t \geq 2$ and, if $t = 2$, then $z \neq 0$. 
Moreover, let $m$ be the number of distinct propositional variables among $q_1,\dots,q_t$. 
Then, there exists a sequence $(r_1,\dots,r_n)$ containing all such variables, with $m \leq n$, 
such that the relational sequents $r_1 \ll r_2,\dots,r_{n-1} \ll r_n,r_n \ll r_1$, $\top \ll r_i$ 
and $r_i \preccurlyeq \top$, $i=1,\dots,n$ occur in $G$.
\item[(ii)] Let $\{ p_i : 0 \leq i \leq l \}$ be the propositional variables of $A$. 
Then, the propositional variables occurring in $G$ are exactly $p_1,\dots,p_l$.
\end{enumerate}
\begin{proof} 
($i$) By induction on the depth of $T_A$ nodes, we prove that, for all $u \in V$, if the relation $A_1,\dots,A_s \triangleleft A_{s+1},\dots,A_t$ occurs in the label $H$ of $u$ and $m$ is the number of distinct formulas among $A_1,\dots,A_t$, 
then, for some sequence $(B_1,\dots,B_n)$ containing all such formulas, with $m \leq n$, the relations $B_1 \ll B_2,\dots,B_{n-1} \ll B_n$, $\top \ll B_i$ and $B_i \preccurlyeq \top$, $i=1,\dots,n$, occur in $H$. 
The required conclusion follows immediately from the fact that $G$ is irreducible, that is, for $i=1,\dots,t$, $A_i$ is a propositional variable. 

For the base case, if $d(u)=0$, $u$ is the root of $T_A$ and the claim holds trivially since no relations of the required form occur in $u$.

For the inductive step, 
we assume by induction hypothesis that, for all nodes $u^\prime \in V$ with $d(u^\prime)=i$, 
if the relation $A^\prime_1,\dots,A^\prime_{s^\prime} \triangleleft A^\prime_{s^\prime+1},\dots,A^\prime_{t^\prime}$ occurs in the label $H^\prime$ of $u^\prime$ 
and $m^\prime$ is the number of distinct formulas among $A_1,\dots,A_{t^\prime}$, 
then, for some sequence $(B^\prime_1,\dots,B^\prime_{n^\prime})$ containing all such formulas with $m^\prime \leq n^\prime$,
the relations $B^\prime_1 \ll B^\prime_2,\dots,B^\prime_{n^\prime-1} \ll B^\prime_{n^\prime}$, 
$\top \ll B^\prime_i$ and $B^\prime_i \preccurlyeq \top$, $i=1,\dots,n^\prime$, occur in $H^\prime$. 
We prove that, for any node $u$ of depth $i+1$ such that the relation $A_1,\dots,A_s \triangleleft A_{s+1},\dots,A_t$ occurs in the label $H$ of $u$ 
and $m$ is the number of distinct formulas among $A_1,\dots,A_{t}$, 
then, for some sequence $(B_1,\dots,B_n)$ containing all such formulas with $m \leq n$, 
the relations $B_1 \ll B_2,\dots,B_{n-1} \ll B_n$, $\top \ll B_i$ and $B_i \preccurlyeq \top$, $i=1,\dots,n$, occur in $H$. 
We examine two cases. 

In the first case, $H$ is a premise of the rewriting rule for $\odot$ with conclusion $H^\prime$ and a pivot $C \odot D$. 
If $H$ is the $\odot_1$, $\odot_2$ or $\odot_5$ premise, the claim follows directly from the induction hypothesis. 
If $H$ is the $\odot_3$ premise, then we study the antecedent and the consequent separately. 
For the antecedent, 
we observe that a relation of the form $\triangleleft$, $\triangleleft \in \{ \prec_z, \preccurlyeq_z \}$, occurs in the antecedent, 
namely $C,D \prec_{-1}$, but, $C \ll D$, $D \ll C$, $\top \ll C$, $\top \ll D$, $C \preccurlyeq \top$ and $D \preccurlyeq \top$ occur in the antecedent as well. 
For the consequent, we observe that the formula $C \odot D$ is replaced by the pair $C,D$ in $\triangleleft$ relations 
and by the less complex among $C,D$, say $C$, in $\ll$ relations, 
thus, on the one hand, the cycle we are assuming by induction hypothesis is preserved with $C$ instead of $C \odot D$ 
and moreover it includes also $D$, since we know that both $C \ll D$ and $D \ll C$ occur the antecedent. 
On the other hand, the $\triangleleft$ relations in which $C,D$ occur are covered since also $C \preccurlyeq \top$ and $D \preccurlyeq \top$ occur the antecedent. 
If $H$ is the $\odot_4$ premise, the argument is similar.

In the second case, $H$ is a premise of the rewriting rule for $\rightarrow$ with conclusion $H^\prime$ and a pivot $C \rightarrow D$. 
If $H$ is the $\rightarrow_1$ or $\rightarrow_3$ premise, the claim follows directly from the induction hypothesis. 
If $H$ is the $\rightarrow_2$, the argument is similar to the previous carried out for $\odot_{3,4}$ premises.

($ii$) By induction on the depth of $T_A$ nodes, we prove that, for all $u \in V$ with label $H$, the propositional variables occurring in $H$ are exactly the propositional variables of $A$. 

For the base case, if $d(u)=0$, $u$ is the root of $T_A$ and the claim holds trivially. 
For the inductive step, we assume by induction hypothesis that the claim holds for all nodes $u^\prime$ with label $H^\prime$ of depth $i$ 
and we prove that the claim holds for a node $u$ with label $H$ of depth $i+1$. 
If $H$ is a premise $\odot_1$, $\odot_2$, $\odot_3$ or $\odot_4$ of the rewriting rule for $\odot$, 
or also a premise $\rightarrow_1$ or $\rightarrow_2$ of the rewriting rule for $\rightarrow$, 
all the variables of the pivot occur in the antecedent of the premise. 
The other variables are unaffected by the rule and we can apply the induction hypothesis. 
If $H$ is a premise $\odot_5$ or $\rightarrow_3$, all the variables of the pivot occur in the antecedent of the premise. 
However, by Definition~\ref{RHBLRewritingRules} the $\prec_z,\preccurlyeq_z$ relations (with more than one formula on the right or left side or with $z \neq 0$) have been removed from $H$ together with their propositional variables. 
But the previous claim ($i$) guarantees that the variables removed occur in $\ll$ relations in $H$.
The other variables are unaffected by the rule and we can apply the induction hypothesis.
\end{proof}
\end{proposition}

\section{Axiom Check and Countermodel Building} \label{axiomSection}

The main goal of this section is to show that 
the problem of checking if the leaf of a \textbf{RWBL} reduction is an axiom or not 
can be decided by a suitable algorithm in time polynomial in the size of the input. 
We refer the reader to 
\cite{CLRS} and \cite{schrijver:theory}
for standard notions and facts in the fields of 
algorithms and linear programming respectively.

Preliminarily, we formalize the notion of size for formulas and relational hypersequents. 
We will assume a finite alphabet $\Sigma$ and a standard encoding $\langle \cdot \rangle$ 
such that, for all $s \in \Sigma^*$, $\langle s \rangle \in \{ 0,1\}^*$ and $|\langle s \rangle| = O(|s|)$ 
For definiteness, we will assume that the character set $\Sigma$ and the character encoding $\langle \cdot \rangle$ 
are provided by the ASCII standard. From now on, formulas and relational hypersequents will be considered as strings over $\Sigma$, that is, $F,RH \subseteq \Sigma^*$. 
Let $A$ be a formula with $n_0$ occurrences of connectives and $n_1$ occurrences of variables. 
Avoiding redundant parenthesis, it is easy to define the syntax of formulas providing that $|A|=O(n_0)$, since $n_1=n_0+1$. 
Let $G$ be a relational hypersequent with $n_0$ occurrences of connectives, 
$n_1$ occurrences of variables and $n_2$ occurrences of relations. 
Observing that variables and relations bring at most a constant number of parenthesis and commas, 
it is easy to define the syntax of relational hypersequents providing that $|G|=O(n_0+n_1+n_2)$. 

The following definition is legitimate in the light of the previous discussion.

\begin{definition}[size] 
Let $A \in F$ with $n_0$ occurrences of connectives. 
The \emph{size} of $A$ is the length $|\langle A \rangle|$ of its standard binary encoding $\langle A \rangle$. 

Let $G$ be a relational hypersequent with $n_0$ occurrences of connectives, $n_1$ occurrences of variables and $n_2$ occurrences of relations. 
The \emph{size} of $G$ is the length $|\langle G \rangle|$ of its standard binary encoding $\langle G \rangle$. 
\end{definition}

Note that $|\langle A \rangle|=O(n_0)$ and $|\langle G \rangle|=O(n_0+n_1+n_2)$.
It is meant that, with respect to reduction trees, 
the size of a node is the size of the relational hypersequent labelling the node, 
the size of a branch is the sum of the sizes of the nodes of the branch and 
the size of the tree is the sum of the sizes of the nodes of the tree.  

We introduce the notion of axiom in the context of \textbf{RWBL} reductions and, 
in the usual formal languages framework, 
the axiom decision problem for a \textbf{RWBL} reduction leaf.

\begin{definition}[axiom, BL-AX] 
Let $H$ be the label of a leaf of a \emph{\bf{RWBL}} reduction. Then, $H$ is an \emph{axiom} if it is valid. 
$\emph{BL-AX}=\{ \langle H \rangle : H \text{ is an axiom}\} \subseteq \{ 0,1 \}^*$.
\end{definition}

We consider the problem of checking if the leaf of a \textbf{RWBL} reduction is an axiom or not. 
An instance of the problem BL-AX is represented by the leaf of a \textbf{RWBL} reduction, 
that is, an irreducible relational hypersequent $H$ of the form $H_1|\dots|H_k$ 
corresponding to the boolean disjunction $H_1 \vee \dots \vee H_k$. The question is if $H$ is an axiom or not, 
equivalently, if $H$ is valid or not. 
We reduce to the question if the negation $\overline{H}$, 
of the form $\neg H_1 \wedge \dots \wedge \neg H_k$ is satisfiable or not. 
If $\overline{H}$ is satisfiable, then $H$ is not an axiom, 
otherwise, if $\overline{H}$ is unsatisfiable, $H$ is an axiom.

Let $Var(H)=\{ p_1,\ldots, p_n\}$ be the propositional variables of $H$ and 
$v$ be a valuation. 
Moreover, let $q,r \in Var(H) \cup \{ \top \}$ and $q_1,\dots,q_l,r_1,\dots,r_m \in Var(H)$. 
Then any relational sequent $\neg I \in \{ \neg H_1,\dots, \neg H_k\}$ has one of the following forms
\begin{gather}
\overline{q \ll r} \label{F1} \\
\overline{q \prec r} \label{F2} \\
\overline{q \preccurlyeq r} \label{F3} \\
\overline{q_1,\dots,q_l \prec_z r_1,\dots,r_m} \label{F4} \\
\overline{q_1,\dots,q_l \preccurlyeq_z r_1,\dots,r_m} \label{F5}
\end{gather}
which are respectively satisfied by $v$ only if
\begin{gather}
\lfloor v(q) \rfloor \neq \lfloor v(r) \rfloor \Rightarrow \lfloor v(r) \rfloor < \lfloor v(q) \rfloor \label{INT1} \\
\lfloor v(q) \rfloor=\lfloor v(r) \rfloor \Rightarrow v(r) \leq v(q) \label{INT2} \\
\lfloor v(q) \rfloor=\lfloor v(r) \rfloor \Rightarrow v(r) < v(q) \label{INT3} \\
\lfloor v(q_1) \rfloor = \cdots = \lfloor v(r_m) \rfloor < +\infty \Rightarrow
\sum_{i=1}^{l}-\overline{v}(q_i)+\sum_{i=1}^{m}\overline{v}(r_i) \leq m-l-z \label{INT4} \\
\lfloor v(q_1) \rfloor = \cdots = \lfloor v(r_m) \rfloor < +\infty \Rightarrow
\sum_{i=1}^{l}-\overline{v}(q_i)+\sum_{i=1}^{m}\overline{v}(r_i) < m-l-z \label{INT5}
\end{gather}

We describe a three stages algorithm for BL-AX, called \textsc{CheckAxiom}. 

\paragraph{Stage 1.} The first stage of the algorithm 
is devoted to the initialization of a directed vertex labeled graph $G=(V,E)$. 
First, $V=\{ v_1,\dots,v_n,v_{n+1} \}$, $n=|Var(H)|$, and the label of a vertex $v$ is $l(v) \subseteq Var(H) \cup \{ \top \}$. 
We identify a vertex by its label. 
At initialization time, for $i=1,\dots,n$, $l(v_i)=\{ p_i \}$ and $l(v_{n+1})=\{ \top \}$. 
Moreover, for each $\neg I \in \{ \neg H_1,\dots, \neg H_k\}$ of the form (\ref{F1}), the set $E$ contains and edge $(r,q)$. 

\paragraph{Stage 2.} The second stage of the algorithm 
is devoted to define the constraints of a linear program $P$ 
of the form $Ax < b, Bx \leq c$, where
\begin{equation*} \label{LP}
x \in \emph{\bf{R}}^{n \times 1},
A \in \emph{\bf{Z}}^{(n_1+n) \times n}, 
b \in \emph{\bf{Z}}^{(n_1+n) \times 1}, 
B \in \emph{\bf{Z}}^{(n_2+n) \times n}, 
c \in \emph{\bf{Z}}^{(n_2+n) \times 1}
\end{equation*}
and $0 \leq n_1,n_2 \leq k$, where $n_1,n_2$ are at most equal to the number of $H_i$s 
of the form (\ref{F3}),(\ref{F5}) and (\ref{F2}),(\ref{F4}) respectively.

To this aim, the cycles eventually contained in $G$ are iteratively detected and removed, 
until the graph $G$ becomes a forest. Specifically, 
if a cycle $c$ is detected, 
say $c=\langle u_1,\dots,u_m,u_1 \rangle$, $m \leq n+1$, 
the vertices $u_1,\dots,u_m$ are removed from $V$ and a vertex $v$, 
with $l(v)=\bigcup_{i=1}^m l(u_i)$, is added to $V$. 
All the edges $(u_i,u_j)$ with $i,j \in \{ 1,\dots,m \}$ are removed from $E$. 
Moreover, 
whenever $u_i$ is in the cycle, $u_j$ is not in the cycle and $(u_i,u_j) \in E$ (respectively $(u_j,u_i) \in E$), 
$(u_i,u_j)$ is removed and replaced by $(v,u_j)$ (respectively $(u_i,v)$). 
At each iteration, $|V|$ decreases at least by one, then the loop iterates at most $n$ times. 
At the end of the loop, the graph $G$ is topologically sortable 
to obtain an ordered list $v_0,\dots,v_l$ of the vertices such that, 
if $G$ contains ad edge $(v_i,v_j)$, then $v_i$ appears before $v_j$ in the ordering. 

Depending on the form of each $\neg I \in \{ \neg H_1,\dots, \neg H_k\}$, 
the constraints of $P$ are initialized.
\begin{enumerate}
\item[$(i)$] If $\neg I$ is like (\ref{F1}), no constraints are added to $P$. 
\item[$(ii)$] If $\neg I$ is like (\ref{F2}), there are two cases. 
If $q=p_i$ and $r=p_j$ for some $i,j \in \{ 1,\dots,n\}$ 
and, for some $m \in \{ 0,\dots,l \}$, both $p_i,p_j \in l(v_m)$, 
then the constraint $x_j \leq x_i$ is added to $P$ 
(no constraints are added if such $m$ does not exist). 
Otherwise, if $q=\top$ or $r=\top$ and, for some $m \in \{ 0,\dots,l \}$, both $q,r \in l(v_m)$, 
the constraint $0 \leq 0$ is added to $P$ 
(no constraints are added if such $m$ does not exist). 
\item[$(iii)$] If $\neg I$ is like (\ref{F3}), there are two cases. 
If $q=p_i$ and $r=p_j$ for some $i,j \in \{ 1,\dots,n\}$ 
and, for some $m \in \{ 0,\dots,l \}$, both $p_i,p_j \in l(v_m)$, 
then the constraint $x_j < x_i$ is added to $P$ 
(no constraints are added if such $m$ does not exist). 
Otherwise, if $q=\top$ or $r=\top$ and, for some $m \in \{ 0,\dots,l \}$, both $q,r \in l(v_m)$, 
then the constraint $0 < 0$ is added to $P$ 
(no constraints are added if such $m$ does not exist). 
\item[$(iv)$] If $\neg I$ is like (\ref{F4}), there are two cases. 
If there exist $m,m^{\prime} \in \{ 0,\dots,l \}$ such that 
all the variables of $\neg I$ occur in $l(v_m)$, $\top \notin l(v_m)$, $\top \in l(v_{m^{\prime}})$, $(v_m,v_{m^{\prime}}) \in E$ and $(v_{m^{\prime}},v_m) \notin E$, 
then $P$ is extended with the constraint $b_1 \cdot x_1 + \cdots + b_n \cdot x_n \leq c$, 
where $b_1,\dots,b_n,c$ are the integers derived from (\ref{INT4}), 
otherwise, no constraints are added to $P$.
\item[$(v)$] If $\neg I$ is like (\ref{F5}), there are two cases. 
If there exist $m,m^{\prime} \in \{ 0,\dots,l \}$ such that 
all the variables of $\neg I$ occur in $l(v_m)$, $\top \notin l(v_m)$, $\top \in l(v_{m^{\prime}})$, $(v_m,v_{m^{\prime}}) \in E$ and $(v_{m^{\prime}},v_m) \notin E$, 
then $P$ is extended with the constraint $a_1 \cdot x_1 + \cdots + a_n \cdot x_n < b$, 
where $a_1,\dots,a_n,b$ are the integers derived from (\ref{INT5}), 
otherwise, no constraints are added to $P$. 
\end{enumerate}
Finally, the constraint $0 \leq x_1,\dots,x_n < 1$ is added to $P$.

\paragraph{Stage 3.} The third stage of the algorithm 
is devoted to check if the linear program $P$ is feasible or not. 
If $P$ is feasible, then the algorithm returns in output ``No'', 
otherwise it returns in output ``Yes''.

\begin{example} \label{ExCheckAx2}
Let $H$ be 
$\overline{p_1,p_2 \prec_{-1}}|p_1 \ll p_1|p_1,p_1,p_2 \prec_{-1} p_1,p_2|p_1,p_1,p_2 \preccurlyeq_{-1} p_1,p_2|p_1,p_2 \prec_{1} p_1,p_1,p_2|p_1,p_2 \preccurlyeq_{1} p_1,p_1,p_2|\top \preccurlyeq p_1$. 
By Notation~\ref{NotationAbbreviations}, $H$ is the boolean disjunction of 
$p_1 \ll p_2$, 
$p_2 \ll p_1$, 
$p_1 \preccurlyeq \top$, 
$\top \preccurlyeq p_1$, 
$\top \ll p_1$, 
$p_2 \preccurlyeq \top$, 
$\top \preccurlyeq p_2$, 
$\top \ll p_2$, 
$\preccurlyeq_{1} p_1,p_2$,
$p_1 \ll p_1$, 
$p_1,p_1,p_2 \prec_{-1} p_1,p_2$, 
$p_1,p_1,p_2 \preccurlyeq_{-1} p_1,p_2$, 
$p_1,p_2 \prec_{1} p_1,p_1,p_2$, 
$p_1,p_2 \preccurlyeq_{1} p_1,p_1,p_2$, 
$\top \preccurlyeq p_1$. 

Consequently, $\overline{H}$ is the boolean conjunction of 
$\overline{p_1 \ll p_2}$, $\overline{p_2 \ll p_1}$, 
$\overline{p_1 \preccurlyeq \top}$, $\overline{\top \ll p_1}$, $\overline{\top \preccurlyeq p_1}$, 
$\overline{p_2 \preccurlyeq \top}$, $\overline{\top \ll p_2}$, $\overline{\top \preccurlyeq p_2}$, 
$\overline{\preccurlyeq_{1} p_1,p_2}$, 
$\overline{p_1 \ll p_1}$, 
$\overline{p_1,p_1,p_2 \prec_{-1} p_1,p_2}$, 
$\overline{p_1,p_1,p_2 \preccurlyeq_{-1} p_1,p_2}$, 
$\overline{p_1,p_2 \prec_{1} p_1,p_1,p_2}$, 
$\overline{p_1,p_2 \preccurlyeq_{1} p_1,p_1,p_2}$, 
$\overline{\top \preccurlyeq p_1}$. 

At initialization time 
the graph $G$ has 
vertices $V=\{ u_1, u_2, u_3 \}$, 
labels $l(u_1)=\{ p_1 \}$, $l(u_2)=\{ p_2 \}$, $l(u_3)=\{ \top \}$ and 
edges $p_1 \rightarrow p_1, p_1 \rightarrow p_2$, $p_2 \rightarrow p_1$, $p_1 \rightarrow \top$, $p_2 \rightarrow \top$.
At the end of the clustering loop 
$G$ has 
vertices $V=\{ v_0,v_1 \}$, 
labels $l(v_0)=\{ p_1,p_2 \}$, $l(v_1)=\{ \top \}$ and 
edges $\{ p_1,p_2 \} \rightarrow \top$. 
The topological sorting 
of $G$ is $(v_0,v_1)$. 

At the end of the linear program initialization loop, 
the linear program $P$ has the form $Ax < b, Bx \leq c$, where $x^T=\begin{pmatrix}x_1 \cdots x_2\end{pmatrix} \in \emph{\bf{R}}^{2 \times 1}$ and
{\small 
\begin{gather*}
A=
\begin{pmatrix}
1 & 1\\
1 & 0\\
-1 & 0\\
1 & 0\\
0 & 1\\
\end{pmatrix}
\in \emph{\bf{Z}}^{5 \times 2}, 
b=
\begin{pmatrix}
1\\
0\\
0\\
1\\
1
\end{pmatrix}
\in \emph{\bf{Z}}^{5 \times 1},
B=
\begin{pmatrix}
1 & 0\\
-1 & 0\\
-1 & 0\\
0 & -1
\end{pmatrix}
\in \emph{\bf{Z}}^{4 \times 2}, 
c=
\begin{pmatrix}
0\\
0\\
0\\
0
\end{pmatrix}
\in \emph{\bf{Z}}^{4 \times 1}. 
\end{gather*}}
The linear program $P$ is unfeasible, 
since it contains the constraints $x_1 < 0$ and $0 \leq x_1$, 
then $H$ is an axiom.
\end{example}

To the aim of characterize the soundness of \textsc{CheckAxiom}, 
it is useful to observe that the clustering loop 
maintains the invariant that, 
at each iteration, each propositional variable of $H$ occurs in the label of exactly one vertex of $G$. 
This property holds at initialization time and guarantees, when the loop terminates, 
that the labels of the vertex of $G$ form a partition of the propositional variables of $H$.

\begin{proposition} \label{UnsatCharacterization}
The relational hypersequent $H$ is an axiom if and only if the linear program $P$ is unfeasible.
\begin{proof}
Equivalently, we show that $\overline{H}$ is unsatisfiable if and only if $P$ is unfeasible.

$(\Rightarrow)$ We prove that, if $P$ is feasible and 
$\overline{x}=\begin{pmatrix}\overline{x_1} \cdots \overline{x_n}\end{pmatrix}^T$ is a solution, 
then there is a valuation $v$ such that $v$ satisfies $\neg H_i$ for all $i=1,\dots,k$. 

We define a valuation $v$ such that, for all $i=1,\dots,n$, 
the valuation $v(p_i)$ is the sum of an integer part, obtained from the topological sorting $v_0,\dots,v_m$ of $G$, 
and a decimal part, obtained from the feasible solution $\overline{x}$ of $P$. 
Specifically, if $p_i \in l(v_j)$ and $\top \notin l(v_j)$, $j \in \{ 0,\dots,m\}$, then $v(p_i)=j+\overline{x_i}$, 
that is, $\lfloor v(p_i) \rfloor=j$ and $\overline{v}(p_i)=\overline{x_i}$. 
Otherwise, if $p_i \in l(v_j)$ and $\top \in l(v_j)$, then $v(p_i)=+\infty$. 
Notice that $v$ is well defined, by the invariant of clustering loop. 
We show that $v$ satisfies $\neg H_i$, for all $i=1,\dots,k$.

If $\neg H_i$ has the form $\overline{q \ll r}$, we examine two cases. 
If there is a cycle $c=\langle r,q,\dots,r \rangle$ in $G$ at initialization time, 
then 
both $q,r \in l(v_j)$ for some $j \in \{ 0,\dots,m\}$, 
then $\lfloor v(p) \rfloor = \lfloor v(q) \rfloor = j$ and $v$ satisfies $\neg H_i$. 
Otherwise, 
$r \in l(v_i)$ and $q \in l(v_j)$ for some $i < j$, with $i,j \in \{ 0,\dots,m\}$, 
by the properties of the topological sorting, 
then $\lfloor v(r) \rfloor < \lfloor v(q) \rfloor$ and $v$ satisfies $\neg H_i$. 

If $\neg H_i$ has the form $\overline{q \prec r}$, we examine two cases. 
If there is a cycle $c=\langle r,q,\dots,r \rangle$ in $G$ at initialization time, 
then both $q,r \in l(v_j)$ for some $j \in \{ 0,\dots,m\}$ 
and, reasoning as before, $\lfloor v(p) \rfloor = \lfloor v(q) \rfloor$. 
Now, if $q=p_i$ and $r=p_j$, for some $i,j \in \{ 1,\ldots,n \}$, 
then the constraints $x_j \leq x_i$, $0 \leq x_i,x_j < 1$, have been added to $P$, 
then $v(r) \leq v(q)$ since $\overline{x_j}=\overline{v}(p_j)$ and $\overline{v}(p_i)=\overline{x_i}$. 
Otherwise, if $q=\top$ or $r=\top$, by the definition of $v$, $v(r) \leq v(q)$ holds since $v(r)=v(q)=+\infty$. 
In all cases, $v$ satisfies $\neg H_i$. 
If there is not a cycle $c$ in $G$ at initialization time, 
$r \in l(v_i)$ and $q \in l(v_j)$ for some $i \neq j$, with $i,j \in \{ 0,\dots,m\}$, 
then $\lfloor v(r) \rfloor \neq \lfloor v(q) \rfloor$ and $v$ satisfies $\neg H_i$. 
If $\neg H_i$ has the form $\overline{q \preccurlyeq r}$, 
the argument is similar with the exception that the case in which 
there is a cycle $c=\langle r,q,\dots,r \rangle$ in $G$ at initialization time and $q=\top$ or $r=\top$ 
is excluded (in this case $P$ is unfeasible, since at Stage 2($iii$) the constraint $0 < 0$ is added to $P$). 

If $\neg H_i$ has the form (\ref{F4}), 
by Proposition~\ref{LeafProperty}($i$) there is a cycle $c$ in $G$ at initialization time 
involving all the variables of $\neg H_i$. 
We examine two cases. 
If $\top$ is involved in $c$, then there exists $j \in \{ 0,\dots,m\}$, 
such that all the variables of $\neg H_i$ and $\top$ are in $l(v_j)$, 
but then $v$ satisfies $\neg H_i$, since by definition $v$ does not satisfy the antecedent of (\ref{INT4}). 
Otherwise, if $\top$ is not involved in $c$, 
then by Proposition~\ref{LeafProperty}($i$) 
the conditions for adding to $P$ the constraints $0 \leq x_i,\dots,x_j < 1$ and $b_1 \cdot x_1 + \cdots + b_n \cdot x_n < c$, 
for suitable integers $b_1,\dots,b_n,c$ derived from (\ref{INT4}), are satisfied. 
Hence, $v$ satisfies the consequent of (\ref{INT4}), 
since for each variable $p_i$ in $\neg H_i$, $\overline{x_i}=\overline{v}(p_i)$.
If $\neg H_i$ has the form (\ref{F5}), the argument is similar.

$(\Leftarrow)$ We show that, if $\overline{H}$ is satisfiable, then $P$ is feasible. 
Specifically, if a valuation $v$ satisfies $\overline{H}$, we show that 
$\overline{x}=\begin{pmatrix}\overline{v}(p_1) \cdots \overline{v}(p_n)\end{pmatrix}^T$ 
is a solution of $P$. To this aim, observe that, during the second stage of the algorithm, 
there are two cases in which the linear program $P$ is expanded with 
new constraints corresponding to $\neg H_i$s.

In the first case, $\neg H_i$ has the form (\ref{F2}) or (\ref{F3}) 
and there is $m \in \{ 0,\dots,l \}$ such that $q,r \in l(v_m)$.
In this case, the graph $G$ at initialization time contains a cycle $c=\langle p_1,\ldots,p_m,p_1 \rangle$, 
with $q,r \in \{ p_1,\ldots,p_m \}$. 
Therefore, a valuation $v$ can satisfy $\overline{H}$ only if $\lfloor v(p_1) \rfloor = \cdots = \lfloor v(p_m) \rfloor$. 
But in this case, accordingly with (\ref{INT2}) and (\ref{INT3}), 
the valuation $v$ satisfies $\overline{H}$ only if it satisfies the conditions over the decimal parts of $p_1,\ldots,p_m$, 
and, particularly, over the decimal parts of $q,r$. 
But then, the decimal parts of $v$ satisfies also the constraints added to $P$ for the currently examined case.

In the second case, $\neg H_i$ has the form (\ref{F4}) or (\ref{F5}), 
and there exist $m,m^{\prime} \in \{ 0,\dots,l \}$ such that 
all the variables of $\neg I$ occur in $l(v_m)$, $\top \notin l(v_m)$, $\top \in l(v_{m^{\prime}})$, $(v_m,v_{m^{\prime}}) \in E$ and $(v_{m^{\prime}},v_m) \notin E$. 
In this case, the graph $G$ at initialization time contains a cycle $c$ involving all the variables of $\neg I$, but not $\top$, 
and each vertex labeled by a variable of $\neg I$ is connected to the vertex labeled by $\top$, but the converse does not hold. 
Therefore, a valuation $v$ can satisfy $\overline{H}$ only evaluating each variable with the same integer part. 
Moreover, by Proposition~\ref{LeafProperty}($i$), a valuation $v$ can satisfy $\overline{H}$ only assigning integer parts less than $+\infty$ to the variables, 
for otherwise $v$ must satisfy also $\top \prec x$ for all the variables of $\overline{H}$, which is a contradiction. 
But then, a valuation $v$ can satisfy $\overline{H}$ only if it satisfies also the antecedent of the implications (\ref{INT4}) and (\ref{INT5}), 
hence such a valuation must satisfy the consequent of the implications (\ref{INT4}) and (\ref{INT5}). 
But then, the decimal parts of $v$ satisfies also the constraints added to $P$ for the currently examined case.
\end{proof}
\end{proposition}

%Notice that Proposition~\ref{LeafProperty}($i$), 
%exploited in the the proof of the left direction of Proposition~\ref{UnsatCharacterization}, 
%holds if the irreducible relational hypersequent $H$ is the leaf of a \textbf{RWBL} reduction, 
%but does not hold for arbitrary irreducible relational hypersequents. 
%For this reason, the algorithm \textsc{CheckAxiom} is sound for the problem we are interested in, BL-AX, 
%but it is not sound for the problem of checking the validity of arbitrary irreducible relational hypersequents, 
%which is in fact a different problem.

We conclude our argument proving that the problem BL-AX is in fact in \textbf{P}.

\begin{lemma} \label{AxiomCheckP}
$\emph{BL-AX} \in \emph{\bf{P}}$.
\begin{proof}
The problems of searching graphs for cycles, topologically sort forests and checking the feasibility of linear programs are in \textbf{P}. 
Thus, the termination and soundness of \textsc{CheckAxiom} follow from the finiteness of the constructions of $G$ and $P$ and 
Proposition~\ref{UnsatCharacterization}, respectively, and \textsc{CheckAxiom} is a decision algorithm for BL-AX. 
Moreover, both $|\langle G \rangle|$ and $|\langle P \rangle|$ are polynomial in $|\langle H \rangle|$, 
for any reasonable binary representation of graphs and matrices, 
and the constructions involved can be easily implemented in time polynomial in the size of the objects built, 
then \textsc{CheckAxiom} runs in polynomial time in the size $|\langle H \rangle|$ of its input $\langle H \rangle$. 
We conclude that $\text{BL-AX} \in \emph{\bf{P}}$.
\end{proof}
\end{lemma}

We remark that, if a leaf $H$ is not valid, 
the proof of the right direction of Proposition~\ref{UnsatCharacterization} provides an effective means for constructing a countermodel of $H$, 
which propagates to the root of the reduction tree yielding the following countermodel building procedure.

Let $H \in RH$ be a leaf of the \textbf{RWBL} reduction tree of a formula $A$ and suppose that the valuation $v$ is a countermodel of $H$. 
Let $G_0,\ldots,G_n$ be the labels of the nodes along the branch from the root to the leaf labeled by $H$, 
where $H=G_0$ and $G_n$ is the label of the root. 
For $i=1,\ldots,n$, 
by Proposition~\ref{LeafProperty}($ii$), $v$ is defined on the propositional variables of $G_i$, 
and, by the \emph{invertibility} of the rewriting rules, if $v$ is a countermodel of $H$, then $v$ is a countermodel of $G_i$. 
Particularly, $v$ is a countermodel of the root $\top \preccurlyeq A$, then $v$ satisfies $A \ll \top$. 
Therefore, there is a valuation $v$ such that $v(A) < +\infty$, and, by Theorem~\ref{Montagna}, $\textbf{BL} \nvdash A$.

\section{Decidability and Complexity} \label{conpcSection}

The main goal of this section is to show that the tautology problem of \textbf{BL} is decidable in 
deterministic time $2^{O(n)}$. Also, we show that the problem is \textbf{coNP}-complete. 
In the usual formal languages framework, the tautology decision problem of \textbf{BL} is defined as follows.

\begin{definition}[BL-TAUT] 
$\emph{BL-TAUT}=\{ \langle A \rangle : \emph{\bf{BL}} \vdash A \} \subseteq \{ 0,1 \}^*$.
\end{definition}

We study first the decidability of BL-TAUT, referring to the algorithm \textsc{CheckTautology} 
described as follows. 
The algorithm receives in input (the encoding of) a formula $A$ and builds the \textbf{RWBL} reduction tree of $A$. 
The root of the tree is labeled with the relational hypersequent $\top \preccurlyeq A$. 
The tree is built adding children to reducible external nodes, until all external nodes become irreducible. 
The main connective of the most complex formula occurring in the reducible external node determines the multiplicity and the labels of its children. 
Then, the algorithm checks if all the leaves are axioms and returns a positive answer if they are and a negative answer otherwise. 

In order to prove that the algorithm \textsc{CheckTautology} is in fact a decision algorithm for BL-TAUT, 
we must show that, given a formula $A$, the algorithm \textsc{CheckTautology} always terminates, 
returning a positive answer if $A$ is a \textbf{BL} tautology and a negative answer otherwise.
 
These properties rely respectively on the finiteness of the reductions, 
which is shown in Lemma~\ref{Termination} below, 
and on the soundness and invertibility of the rewriting rules, 
which is shown in Corollary~\ref{RewritingSoundness}. 

\begin{lemma} \label{Termination}
Let $A \in F$ with $c(A)=n$ and $T_A=(V,E)$ be the \emph{\bf{RWBL}} reduction of $A$. 
Then, for all leaves $u$ of $(T_A)$, $d(u) \leq n$, and $h(T_A) \leq n$.
\begin{proof}
Let $A_1,\ldots,A_n$ be the $n$ subformulas of $A$ with at least one connective. 
We assume, without loss of generality, that $A_1 <_c \ldots <_c A_n$. 
We say for short that a node has a subformula 
if the subformula is a subformula of some of the formulas occurring in its label.

By induction on the depth of the nodes, exploiting Definition~\ref{RHBLRewritingRules} and Definition~\ref{RewritingTree}, 
we prove that, for all $u \in V$, if $d(u)=i$, then $u$ has at most $n-i$ 
subformulas of $A$ with at least one connective, for some and $i \leq n$. 
For the base case, if $d(u)=0$, then $u$ is the root. 
Hence, $u$ has $n-0=n$ subformulas of $A$ with at least one connective 
(and $0 \leq n$). 
For the inductive step, we assume that, for all nodes $u^\prime \in V$, 
if $d(u^\prime)=i$, then $u^\prime$ has at most $n-i$ subformulas of $A$ with at least one connective, 
for some $i \leq n$. 
We must prove that, if a node $u$ has depth $i+1$, 
then $u$ has at most $n-(i+1)$ subformulas of $A$ with at least one connective, $i+1 \leq n$. 
If $d(u)=i+1$, then $u$ is a children of a parent $u^\prime$ with $d(u^\prime)=i$ and, by inductive hypothesis, 
$u^\prime$ has at most $n-i$ subformulas of $A$ with at least one connective, $i \leq n$. 
Notice that $i < n$, since $u^\prime$ is not a leaf, thus, by the construction of Definition~\ref{RewritingTree}, 
$u$ has at least subformula of $A$ with at least one connective less than its parent $u^\prime$, 
that is, $u$ has at most $(n-i)-1=n-(i+1)$ subformulas of $A$ with at least one connective, where $i+1 \leq n$. 

Now, consider a leaf $u$ of $(T_A)$ and suppose, for a contradiction, 
that $d(u)=m$, for some $m>n$. 
Then $u$ would have $n-m$ subformulas of $A$ with at least one connective, 
but $0 \leq n-m < 0$, which is a contradiction. 
Thus, for all leaves $u$ of $(T_A)$, $d(u) \leq n$. 
But the height of $T_A$ is equal to the largest depth of any leaf in the tree, 
hence $h(T_A) \leq n$.
\end{proof}
\end{lemma}

We can conclude that \textsc{CheckTautology} decides BL-TAUT.

\begin{theorem}
Let $A$ be a formula. Then, $A \in \emph{\text{BL-TAUT}}$ if and only the algorithm \textsc{CheckTautology} outputs ``Yes'' on input $\langle A \rangle$.
\end{theorem}
Now, we study the complexity of BL-TAUT. 
For this purpose, we show that the overall size of a reduction tree is exponential in the size of the root, 
but the size of each branch is polynomial in the size of the root.

\begin{lemma} \label{Complexity} 
Let $A \in F$ with $c(A)=n$, $T_A=(V,E)$ be the \emph{\bf{RWBL}} reduction of $A$ and $b$ a $T_A$'s branch. 
Then $|\langle b \rangle|=O(n^3)$ and $|\langle T_A\rangle|=2^{O(n)}$.
\begin{proof}
Let $b=(u_0, u_1, \ldots, u_n)$ be a branch of $T_A$ of maximal length 
(recall Lemma~\ref{Termination}), where $u_0$ is the root of $T_A$. 
We must compute the size of the labels $l_0, \ldots, l_n$ of the nodes $u_0, \ldots, u_n$ of $b$. 
For this purpose, we represent the labels by the rows of a lower triangular matrix of relational hypersequents
\begin{equation}
L=
\begin{pmatrix}
a_{1,1}     & \emptyset  & \dots  & \emptyset \\
a_{2,1}     & a_{2,2}     & \dots  & \emptyset \\
\vdots     & \vdots     & \ddots & \vdots \\
a_{n+1,1} & a_{n+1,2} & \dots  & a_{n+1,n+1}
\end{pmatrix}
\in RH^{(n+1) \times (n+1)}
\end{equation}
in the sense that the label $l_0$ of the node $u_0$ is $a_{1,1}$, 
the label $l_1$ of the node $u_1$ is $a_{2,1}|a_{2,2}$, 
and generally the label $l_{i-1}$ of the node $u_{i-1}$ is $a_{i,1}|\dots|a_{i,i}$. 
The entries of the matrix are to be interpreted as follows (recall the terminology of Definition~\ref{RHBLRewritingRules}). 
Consider first the entries along the diagonal. 
The element $a_{1,1}$ is the label of $u_0$, 
the element $a_{2,2}$ is the antecedent of the relational hypersequent labelling $u_1$, %as determined in Definition~\ref{RHBLRewritingRules},
and generally the element $a_{i,i}$, for all $1 < i \leq n+1$, is the antecedent of the relational hypersequents labelling $u_{i-1}$. 
Now, consider the entries below the diagonal. 
The element $a_{2,1}$ is the consequent of the relational hypersequent labelling $u_1$, 
the elements $a_{3,1}|a_{3,2}$ are the consequent of the relational hypersequent labelling $u_2$, 
and generally the elements $a_{i,1}|\dots|a_{i,i-1}$, for all $1 < i \leq n+1$, are the consequent of the relational hypersequents labelling $u_{i-1}$. 
More specifically, 
with respect to the consequent of $u_1$ 
the element $a_{2,1}$ is to be interpreted as the rewriting of the relational hypersequent $a_{1,1}$, 
with respect to the consequent of $u_2$ 
the elements $a_{3,1}$ and $a_{3,2}$ are to be interpreted respectively as the rewriting of the relational hypersequents $a_{2,1}$ and $a_{2,2}$, 
and generally with respect to the consequent of $u_{i-1}$ the elements $a_{i,1},\dots,a_{i,i-1}$, for $1 < i \leq n+1$, 
are to be interpreted respectively as the rewriting of the relational hypersequents $a_{i-1,1},\dots,a_{i-1,i-1}$. 

Recall that the standard encoding $\langle G \rangle$ of a relational hypersequent $G$ has size $|\langle G \rangle|=O(n_0 + n_1 + n_2)$, 
where $n_0,n_1,n_2$ are respectively the occurrences of connectives, variables and relations in $G$. 
We count $\top$ as a variable. 
Clearly, $O(|\langle G_1 \rangle| + \cdots + |\langle G_n \rangle|) = O(|\langle G_1|\dots|G_n \rangle|)$, for $G_1,\ldots,G_n \in RH$. 
Therefore, in order to determine an upper bound on the size of the branch $b$, 
we can inspect the matrix $L$ and count the number of connectives, variables and relations that occur in $b$ in the worst case. 

First, $a_{1,1}$ is $\top \preccurlyeq A$ and  
it has $n$ connectives, $n+2$ variables and $1$ relation. 
Thus, the root contributes $2n+3$ to the sum. 

Second, the largest antecedent of $b$ is the antecedent $\odot_4$ 
built upon the immediate subformulas of $A$ 
(that is, applying the rule for $\odot$ with pivot $A$), 
as can be easily verified by counting the number of connectives, variables and relations 
in the rules antecedents (recall Definition~\ref{RHBLRewritingRules} and Notation~\ref{NotationAbbreviations}). 
Such an antecedent has $6(n-1)$ connectives, $6(n+1)+6$ variables and $\top$'s occurrences and $9$ relations, 
and it contributes $12n+15$ to the sum. 
We assume that all the antecedents of $b$ are sized as the largest one, 
thus each $a_{i,j}$, for all $1 < i=j \leq n+1$ contributes $12n+15$. 

Third, we observe that the size of the consequent is less than the size of the conclusion, 
with the only exception of the consequent $\odot_4$, 
as can be easily checked by counting inspecting Definition~\ref{RHBLRewritingRules}. 
For this reason, we assume that $a_{2,1}$ is the conclusion $\odot_4$ of the rewriting of $a_{1,1}$, 
thus $a_{2,1}$ has $2(n-1)$ connectives, $2(n+1)$ variables and $\top$'s occurrences and $1$ relation. 
Thus, $a_{2,1}$ contributes $4n+1$ to the sum. 
In the following reductions, that is in $a_{i,1}$ for $2 < i \leq n+1$, 
the size of the conclusion is preserved, 
by the definition of substitution given before Definition~\ref{RHBLRewritingRules} 
(recall also Example~\ref{Substitution2}).
Thus, each $a_{i,1}$, for $2 < i \leq n+1$, contributes at most $4n+1$ to the sum. 
Moreover, the same critical rewriting applied to the largest antecedent previously treated, 
yields to a conclusion having has at most $2 \cdot (6(n-1)-1)$ connectives, $2 \cdot 6(n+1)+6$ variables and $\top$'s occurrences and $9$ relations, 
but subsequently the size is preserved. 
We assume that all the antecedents yield a critical rewriting, 
thus each consequent $a_{i,j}$, for $2<i \leq n+1$ and $1<j \leq n$ contributes $24n+13$. 

Notice that, in a real case, the rules produce repetitions in the relational hypersequents, 
which are disregarded by the definition of relational hypersequent. 
The contribute in terms of connectives, variables and relations 
of each relational hypersequent occurring in $b$ is summarized by the following matrix
\begin{equation} \label{SizeMatrix}
S=
\begin{pmatrix}
2n+3     & 0  & 0  & \dots & 0 \\
4n+1     & 12n+15     & 0  & \dots & 0 \\
4n+1     & 24n+13     & 12n+15 & \dots & 0 \\
\vdots     & \vdots     & \vdots & \ddots & \vdots \\
4n+1 & 24n+13 & 24n+13  & \dots & 12n+15
\end{pmatrix}
\in \emph{\bf{Z}}^{(n+1) \times (n+1)}
\end{equation}
We evaluate the sum $s=n_0+n_1+n_2$ of all the connectives, variables and relations occurring in the relational hypersequents of $b$ by
\begin{equation*}
\begin{split}
s & =    2n+3 + \sum_{i=1}^{n}4n+1 + \sum_{i=1}^{n}12n+15 + \sum_{i=1}^{n}(24n+13)(n-i) \\
  & \leq (24n+13)\sum_{i=0}^{n}i+1\\
  & \leq (24n+13)(n+1)^2 = 24n^3+61n^2+50n+13
\end{split}
\end{equation*}
where the first inequality holds since $n>0$. 
Therefore, we conclude that $|\langle b \rangle|=|\langle l_0 \rangle| + \cdots + |\langle l_n \rangle|=O(n^3)$ 
and the first claim holds. 

For the second claim, we observe that, in the worst case, all the leaves of $T_A$ have depth $n$, by Lemma~\ref{Termination}, 
and all the internal nodes of $T_A$ have degree $5$, by Definition~\ref{ReductionTree}, 
thus $|V|=\sum_{k=0}^{n} 5^k=5^{n+1}-1=2^{O(n)}$. 
Moreover, inspecting the size matrix (\ref{SizeMatrix}), 
we observe that the largest relational hypersequent $l$ labelling a node $u$ of $T_A$ 
has a total of at most $(24n+13)(n+1)$ connectives, variables and relations, thus $|\langle l \rangle|=O(n^2)$. 
Therefore $|\langle T_A \rangle| = |V| \cdot |\langle l \rangle| = 2^{O(n)}$.
\end{proof}
\end{lemma}

As a consequence of Lemma~\ref{AxiomCheckP} and Lemma~\ref{Complexity} above, 
the running time of \textsc{CheckTautology} is bounded from above in $2^{O(n)}$, 
hence the problem $\text{BL-TAUT}$ is in $\text{\bf{DTIME}}(2^{O(n)})$.

Consider now the problem $\overline{\text{BL-TAUT}}=\{ \langle A \rangle : \emph{\bf{BL}} \nvdash A \}$. 
If a formula $A$ is not a \textbf{BL} theorem, or tautology, then the algorithm \textsc{CheckTautology} outputs ``No'', 
but in this case the \textbf{RWBL} reduction tree of $A$ has at least one branch with an invalid leaf, 
which size is polynomial in the size of $A$ by Lemma~\ref{Complexity}. 
We regard such a branch as a certificate that the formula $A$ is not a \textbf{BL} tautology 
and we describe a two input polynomial time algorithm, called \textsc{CheckNoTautology}, as follows.

The first input is the standard encoding of a formula $A$. 
The second input is the encoding of a certificate $c$, 
corresponding to the instructions for building a branch of the \textbf{RWBL} reduction of $A$. 
Particularly, if $c(A)=n$, then $c=(m_1,\ldots,m_n)$, $m_i \in \{ 0,\ldots,5 \}$, for all $i=1,\ldots,n$, and 
each $m_i$ represents the premise for extending the node of depth $i-1$. 
We stipulate for convenience that, if $m_i=0$, then the node of depth $i-1$ has no children, and, 
if $m_i$ is inconsistent with the node of depth $i-1$, then the algorithm outputs ``No''. 
The algorithm builds the branch of the \textbf{RWBL} reduction of $A$, 
following the instructions in $c$, and 
returns a positive answer if the leaf is not valid and a negative answer otherwise.

\begin{corollary}
$\emph{\text{BL-TAUT}} \in \emph{\text{\bf{coNP}}-complete}$.
\end{corollary}
\begin{proof}
We show that $\overline{\text{BL-TAUT}} \in \emph{\bf{NP}}$. 
We consider the algorithm \textsc{CheckNotTautology}, 
that receives in input a formula $A$, with $c(A)=n$, and a certificate $c$ of the form $(m_1,\ldots,m_n)$, with $m_i \in \{ 0,\ldots,5\}$ for all $i=1,\ldots,n$. 

We observe the following facts. 
First, $|\langle c \rangle|=O(n)$, since $c$ is a sequence of $n$ nonnegative natural numbers, 
thus the size of $c$ is polynomial in the size of $A$, which is $|\langle A \rangle|=O(n)$ for $n=c(A)$. 
Second, the size of each $T_A$ branch is polynomial in the size of $A$, by Lemma~\ref{Complexity} 
and its adjacency list representation of has size $\Theta(|V|+|E|)=\Theta((n+1)+n)=O(n)$. 
Thus, the whole construction of the branch specified by $c$ is feasible polynomial time in the size of $A$, 
since each step of the construction is feasible polynomial time in the size of the node extended.
Third, the validity of $T_A$'s leaf, by Lemma~\ref{AxiomCheckP}, is decided polynomial time in the size of the leaf, 
which in turn is polynomial in the size of $A$. 
Fourth, the adjacency list representation of the branch has size $\Theta(|V|+|E|)=\Theta((n+1)+n)$, thus $O(n)$.
Thus \textsc{CheckNotTautology} terminates in polynomial time in the size of the input $A$.

Moreover, \textsc{CheckNotTautology} is sound for $\overline{\text{BL-TAUT}}$. 
In fact, suppose that $A$ is not a \textbf{BL} tautology and that $c$ corresponds to a branch of the reduction tree of $A$ with an invalid leaf. 
Then, on input $(\langle A \rangle,\langle c \rangle)$, the algorithm \textsc{CheckNotTautology} returns the answer ``Yes'' in output. 
Otherwise, if $A$ is a \textbf{BL} tautology, there is no certificate $c$ corresponding to a branch of the reduction tree of $A$ with an invalid leaf, 
thus, on input $(\langle A \rangle,\langle c \rangle)$, the algorithm \textsc{CheckNotTautology} returns the answer ``No'' in output. 

We show that $\overline{\text{BL-TAUT}}$ is \textbf{NP}-hard. 
Let $\overline{\text{\L UK-TAUT}}$ be the complement of the tautology problem of \L ukasiewicz logic, 
which is known \cite{DBLP:journals/tcs/Mundici87} to be \textbf{NP}-hard. 
It is known \cite{DBLP:journals/apal/BaazH01, DBLP:journals/logcom/MontagnaPT03} also that, for each formula $A$, $A$ is provable in \L ukasiewicz logic if and only if $\neg \neg A$ is provable in \textbf{BL}. 
Thus, the polynomial time algorithm that prefixes a double negation to a formula $A$ is a reduction of $\overline{\text{\L UK-TAUT}}$ to $\overline{\text{BL-TAUT}}$.

We conclude that $\overline{\text{BL-TAUT}} \in \emph{\text{\bf{NP}}}\text{-complete}$, thus $\text{BL-TAUT}$ is \textbf{coNP}-complete.
\end{proof}

\end{document}